\documentclass[nologo,url,11pt,a4paper]{ETHpaper}
\usepackage{amssymb,amsmath} 

\usepackage[utf8]{inputenc}

\usepackage{hyperref}

\usepackage{graphicx}
\usepackage{caption}
\usepackage[tight]{subfigure}

\usepackage{placeins}  
\usepackage{animate}

\usepackage{rotating} 
\usepackage{fixltx2e}
\usepackage{mathtools}
\usepackage{tikz, tkz-graph}

\newcommand{\tsp}[1]{\textsuperscript{#1}}

\definecolor{notblack}{rgb}{0.0,0.2,0.7}

\begin{document}

\title{The Network of Counterparty Risk: \\ Analysing Correlations in OTC Derivatives}
\titlealternative{The Network of Counterparty Risk: Analysing Correlations in OTC Derivatives, \\ published in \href{http://dx.doi.org/10.1371\%2Fjournal.pone.0136638}{PLoS ONE 10(9), e0136638. doi: 10.1371/journal.pone.0136638}}
\author{Vahan Nanumyan, Antonios Garas, Frank Schweitzer\footnote{Corresponding author: \texttt{fschweitzer@ethz.ch}}}
\authoralternative{Vahan Nanumyan, Antonios Garas, Frank Schweitzer}
\address{Chair of Systems Design, ETH Zurich, Weinbergstrasse 58, 8092
Zurich, Switzerland}
\www{\url{http://www.sg.ethz.ch}}

\makeframing
\maketitle


\begin{abstract}
Counterparty risk denotes the risk that a party defaults in a bilateral contract. This risk not only depends on the two parties involved, but also on the risk from various other contracts each of these parties holds. In rather informal markets, such as the OTC (over-the-counter) derivative market, institutions only report their \emph{aggregated} quarterly risk exposure, but no details about their counterparties. Hence, little is known about the diversification of counterparty risk. In this paper, we reconstruct the weighted and time-dependent network of counterparty risk in the OTC derivatives market of the United States between 1998 and 2012. To proxy unknown bilateral exposures, we first study the co-occurrence patterns of institutions based on their quarterly activity and ranking in the official report. The network obtained this way is further analysed by a weighted k-core decomposition, to reveal a core-periphery structure. This allows us to compare the activity-based ranking with a topology-based ranking, to identify the most important institutions and their mutual dependencies. We also analyse correlations in these activities, to show strong similarities in the behavior of the core institutions. Our analysis clearly demonstrates the clustering of counterparty risk in a small set of about a dozen US banks. This not only increases the default risk of the central institutions, but also the default risk of peripheral institutions which have contracts with the central ones. Hence, all institutions indirectly have to bear (part of) the counterparty risk of all others, which needs to be better reflected in the price of OTC derivatives. 
\end{abstract}

\section*{Introduction} 
\label{sec:intro}

After the financial crisis of 2008 the systemic risk resulting from OTC (over-the-counter) derivatives has become an important topic of public debate and scientific research. Different from exchange-traded derivatives, OTC derivatives are traded on  non-regulated markets which have grown both in size and importance during the last decade. 
In December 2008 the Bank for International Settlements (BIS) reported (see Semiannual OTC derivatives statistics at \url{http://www.bis.org}) that total notional amount on outstanding OTC derivatives grew up from 370,178 bn USD in June 2006 to 683,725 bn USD in June 2008, i.e., it almost doubled in size in only two years. 

A particular worrying feature  of this development results from the increasing concentration of the counterparty risk of OTC  derivatives in the hands of only a few institutions. 
This trend has not changed after the financial crisis of 2008, on the contrary the concentration increased. 

Taking the example of the US alone, in the 4th quarter of 1998 contracts totaling 331 bn USD were signed by 422 commercial banks and trust companies which where not listed in the top 25 institutions dealing with OTC derivatives. This numbers have to be compared against the contracts totaling 32,668 bn USD (i.e., a hundred times more) signed by only the top 25 institutions in the OTC derivatives market. Comparing this to the time after the financial crisis, the difference became much bigger. 
In the 1st quarter of 2012 the 25 top ranked US institutions held contracts totaling 227,486 bn USD (i.e., almost ten times more than in 1998), whereas all other institutions held contracts totaling only 496 bn USD (which is almost comparable to what was held in 1998). Hence, we observe an extreme concentration of derivatives market where the share of derivative contracts held by the top 25 institutions was almost 99\% in 1998 and increased to more than 99.5\% in 2012. 

This increasing concentration may also increase the vulnerability of the institutions involved and can lead to cascades in case of default.   Until now, no concentration of exposure against a particular counterparty is reported by banks. The Basel Committee on Banking Supervision referred to this issue for the first time  only in its report of March 2013 \cite{BIS}. 

In our paper, we address the problem in a twofold way. Based on a dataset of the 25 most active players in the U.S. derivative market, over a period of 14 years, we reconstruct the network of counterparty risk. We show that this risk generates an almost fully connected network of
interdependence among these players, however it is skewly distributed, i.e., most of the counterparty risk is concentrated in only 10 mayor institutions. 
This implies two problems: in a fully connected network, it becomes much more difficult to hedge the risk of default, because every player is a counterparty of any other. This may increase the risk of default cascades, which can be amplified by the particularly active counterparties.  Additionally, the concentration of counterparty risk in a few institutions may exacerbate the problem of contagion and financial distress in the whole network if those institutions become distressed.

\section*{OTC Derivatives}
\label{sec:otc}

\subsection*{The role of derivatives}
\label{sec:derivatives}

Derivatives are financial instruments, i.e., they are tradable assets. Importantly, they have no intrinsic value. Instead, their value depends on, or is \emph{derived} at least partly from,  the value of other entities, denoted as the ``underlying''. These can be other assets such as commodities, stocks, bonds, interest rates and currencies, but, dependent on the complexity of the financial product, the underlying can be almost anything that deemed to have an intrinsic value. This implies that socio-psychological issues such as ``confidence'', ``faith'' or ``trust'' play an important role in defining those values.

Formally, derivatives are specified as contracts between two parties. Such contracts define how the value of the underlying is estimated at particular future dates and what conditions have to be fulfilled for payments between these parties. Because parties do not need to own the underlying, derivatives make for an ideal instrument to speculate about the future rising or falling value of underlyings or to hedge against the risk associated with it, provided that a counterparty is willing to bet on this.

Trading derivatives basically means to find a counterparty for the contract. Importantly, parties can trade derivatives in two different ways, in regulated markets specialized in trading derivatives (ETD, exchange-traded derivatives) or privately, without involving an exchange or other institutions (OTC, over-the-counter derivatives). 
Although OTC markets are usually well organized, they are less formal. In particular, there is no central authority which would regulate the conditions of the derivative  contracts or would control the fulfillment of these conditions.

OTC derivatives are usually preferred over the exchange traded ones because taxes and other expenses are lower and they are much more flexible, meaning that the counterparties can agree on very specific or unusual conditions as opposed to the limited set of derivative types designed and operated by an exchange. 
As a trade-off for flexibility and the possibility of higher earnings OTC derivatives bear significant additional risks as compared to the exchange traded ones.

\subsection*{Risk involved in OTC derivatives}
\label{otc:risk}

Derivatives are generally used to hedge risks, but derivatives themselves are a source of risk. These are credit risk and market risk, along with liquidity, operational and legal risks \cite{AW}. In case of OTC derivatives, credit risk is the main source of risk because of the usual absence of a clearing house that guarantees the fulfillment of obligations between parties. Thus, the two contracting parties are exposed to \emph{counterparty default risk}, i.e., the risk that a counterparty  will undergo distress, or even default prior to expiration of the contract and thus will not make the current and future payments. In contrast to lending risk, to which only the party which lends is exposed, both sides involved in OTC contract are exposed to counterparty risk. To have some sort of mitigation, the parties involved in OTC derivatives are usually banks which act on their own behalf or on behalf of their clients. 

There are different ways to mitigate counterparty risk in case of default. 
For example,  using close-out netting agreements allows that all contracts are netted, eliminating the possibility of selective execution of contracts \cite{DH}.
For \emph{bilateral close-out netting}, which mostly applies to OTC derivatives markets, the two parties agree to net with one another, i.e., to set off gains and losses from \emph{all} of their bilateral contracts. 
This differs from the case of multilateral close-out netting which mostly applies to ETD, i.e., to markets where all parties' obligations are netted together.
In both cases, netting is only a procedure to follow after a default and thus does not address the \emph{emergence} of counterparty risk.  

It is obvious that netting decreases credit exposure, as it takes into account only the \emph{net} obligations, thus reducing both operational and settlement risk and operational costs. 
In order to know the risk, the \emph{present value} of contracts, prior to their contracted termination, has to be determined. 
Outstanding contracts are marked to market, taking into account the \emph{replacement costs}, i.e., the loss suffered by the non-defaulting party in replacing the relevant contract.
  This assessment of credit exposure at a single point in time is denoted as \emph{current credit exposure} (CCE). 
However, derivative contracts usually have considerable lifetimes and are very often characterized by fast and large changes in credit exposure. 
Therefore, the \emph{potential future exposure} (PFE) is used to estimate the possible CCE increase over a fixed time frame. 
These estimates are, of course, predictions that depend on the choice of financial models and corresponding confidence level.
The \emph{total credit exposure} (TCE) is then measured as the sum of CCE and PFE, following the Basel I framework.
In Section \nameref{sec:corr-risk} we will use the TCE values reported by financial institutions to estimate \emph{correlations} in their risk.

Whereas netting agreements work in the absence of clearing houses, recent developments try to mitigate counterparty risk by means of \emph{central counterparty clearing houses} (CCPs) \cite{Cecchetti2009}.
In the presence of a CCP a bilateral contract between two counterparties is substituted by \emph{two} contracts, so that the CCP stands between the two contracting parties. 
This allows for more transparency and for multilateral netting, which can facilitate the reduction of both counterparty and systemic risk.
Although involvement of a CCP was previously required in contracts for \emph{credit default swaps} (CDS) \cite{Kaushik2013}, a special class of derivatives, 
its broader utilization can be seen as a reaction to the financial crisis of 2008.

However, regulations requiring CCPs in all standardized types of OTC derivatives are either new, e.g.\ the US Dodd-Frank Act from 2010, or are still being developed.
Therefore, their impact on OTC derivatives markets is not well known yet, both empirically and theoretically. \cite{Feng2014} recently attempted to shed some light on the possible systemic effects from CCPs.
They performed a theoretical investigation of cascading effects and systemic risk in different financial networks with one or two CCPs. 

One may argue that 
not considering the role of CCPs in OTC derivatives networks is a limitation of this paper. 
But one should bear in mind that we analyse data ranging from 1998 to 2012, i.e., most of the time CCP were not required, and not reflected, in the OTC data. 
To keep our methods consistent for the whole time period, we neglect the possible (but not documented) presence of CCPs.
Moreover, even today it is not known whether  the wide adoption of CCPs will succeed in making the OTC derivatives network entirely transparent. 
So our methods to infer undiscovered and potentially dangerous links of the network may still be needed in the future.

\subsection*{Clustering of counterparty risk}
\label{sec:clustering}

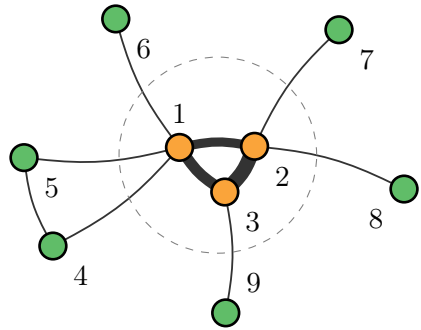
\begin{figure}[htbp]
  \begin{center}
    \definecolor{COLOR0}{rgb}{0.37647058823,0.74117647058,0.40784313725}
\definecolor{COLOR1}{rgb}{0.0,0.0,0.0}
\definecolor{COLOR2}{rgb}{0.98039215686,0.6431372549,0.22745098039}
\definecolor{COLOR3}{rgb}{0.2,0.2,0.2}
\pgfdeclarelayer{background}
\pgfdeclarelayer{foreground}
\pgfsetlayers{background,main,foreground}
\begin{tikzpicture}
\node at (0,3.7169327) [circle, line width=1.0, draw=COLOR1, fill=COLOR0, inner sep=0pt, minimum size = 10.001464pt, label={[label distance=0]315:5}] (8) {};
\node at (0.3799663,2.5452333) [circle, line width=1.0, draw=COLOR1, fill=COLOR0, inner sep=0pt, minimum size = 10.001464pt, label={[label distance=0]315:4}] (5) {};
\node at (5,3.3007294) [circle, line width=1.0, draw=COLOR1, fill=COLOR0, inner sep=0pt, minimum size = 10.001464pt, label={[label distance=0]225:8}] (4) {};
\node at (1.2068703,5.5558776) [circle, line width=1.0, draw=COLOR1, fill=COLOR0, inner sep=0pt, minimum size = 10.001464pt, label={[label distance=0]315:6}] (6) {};
\node at (4.1442257,5.4122805) [circle, line width=1.0, draw=COLOR1, fill=COLOR0, inner sep=0pt, minimum size = 10.001464pt, label={[label distance=0]315:7}] (7) {};
\node at (2.6417727,3.2610484) [circle, line width=1.0, draw=COLOR1, fill=COLOR2, inner sep=0pt, minimum size = 10.001464pt, label={[label distance=0]315:3}] (0) {};
\node at (2.0458545,3.852496) [circle, line width=1.0, draw=COLOR1, fill=COLOR2, inner sep=0pt, minimum size = 10.001464pt, label={[label distance=0]90:1}] (2) {};
\node at (2.653896,1.658776) [circle, line width=1.0, draw=COLOR1, fill=COLOR0, inner sep=0pt, minimum size = 10.001464pt, label={[label distance=0]45:9}] (3) {};
\node at (3.0331331,3.8674649) [circle, line width=1.0, draw=COLOR1, fill=COLOR2, inner sep=0pt, minimum size = 10.001464pt, label={[label distance=0]315:2}] (1) {};
\begin{pgfonlayer}{background}
\tikzset{EdgeStyle/.style = {->,line cap =rect, shorten >=0pt,>=stealth, bend right=10}}
\tikzset{EdgeStyle/.style = {-,line cap =rect, shorten >=0pt, >=stealth, bend right=10, line width=0.6913951, color=COLOR3}}
\Edge (7)(1)
\Edge (4)(1)
\tikzset{EdgeStyle/.style = {-,line cap =rect, shorten >=0pt, >=stealth, bend right=10, line width=4.1483705, color=COLOR3}}
\Edge (2)(0)
\tikzset{EdgeStyle/.style = {-,line cap =rect, shorten >=0pt, >=stealth, bend right=10, line width=3.4569754, color=COLOR3}}
\Edge (1)(2)
\tikzset{EdgeStyle/.style = {-,line cap =rect, shorten >=0pt, >=stealth, bend right=10, line width=0.6913951, color=COLOR3}}
\Edge (8)(5)
\Edge (3)(0)
\Edge (5)(2)
\Edge (6)(2)
\tikzset{EdgeStyle/.style = {-,line cap =rect, shorten >=0pt, >=stealth, bend right=10, line width=5.5311607, color=COLOR3}}
\Edge (0)(1)
\tikzset{EdgeStyle/.style = {-,line cap =rect, shorten >=0pt, >=stealth, bend right=10, line width=0.6913951, color=COLOR3}}
\Edge (8)(2)

\draw[dashed, color=gray] (2.55, 3.75) circle(1.3);
\end{pgfonlayer}
\end{tikzpicture}
  \end{center}
  \caption{Schematic illustration of the exposure clustering.}
  \label{fig:expcluster}
\end{figure}

In this paper, we discuss a particular risk involved in OTC derivatives, namely the \emph{clustering of counterparty risk}. While counterparty risk itself is already difficult to estimate, it becomes even more tedious for a party to find out about the \emph{additional} risk that a counterparty bears because of it's involvement in \emph{other} OTC derivatives. The problem is illustrated in Fig.~\ref{fig:expcluster}. It shows nine institutions that have in total ten different OTC contracts. The width of the links shall indicate the volume of these contracts, i.e., the three institutions 1, 2, 3 in the center (indicated by the dashed line) form a fully connected cluster of strongly engaged institutions. What is their implicit impact on those institutions outside the center? Each of these has only one contract with one of the major institutions in the center and is likely not aware of the whole structure of the network of OTC derivatives. 

There is a two-step scenario to increase the risk of the different institutions: (i) \emph{Transfer of risk from the outer institutions to the central counterparty:} Institution 4 is probably not aware that its counterparty 1 also has contracts with institutions 5 and 6. If one of these outer institutions defaults, this puts an additional risk for institution 1 to default, which is likely not accounted for in the OTC contract between 4 and 1. Additionally, institutions 4 and 5 also have a contract which is likely not known to institution 1. Thus, the default of \emph{either} 4 or 5 increases the risk for the remaining one, which indirectly increases the risk for  institution 1 \cite{Tasca2014}. (ii)  \emph{Increase of risk between central institutions:} Because the center institutions form a fully connected cluster, 
if one of these undergoes distress or even defaults this immediately affects the other two core institutions. This in turn affects the outer institutions. 

In conclusion,  
because of the strong coupling of the center institutions, which we call \emph{clustering of counterparty risk} here, all  institutions indirectly have to bear (part of) the counterparty risk of all other institutions in the network. This should be priced in their OTC derivatives, but effectively it is not because that would imply to  know  (a) all the links and (b) all their weights or, in plain words, all the OTC contracts made. But, as explained above, the existence of OTC derivatives is precisely because such information should \emph{not} be made publicly available. As we will see from the data, all public information only refers to the total amount of OTC derivatives for each institution, but not to their counterparty network.

This sets the stage for our paper. Even in the absence of official information about the network of counterparty risk, we want to derive some insights into its structure, from a dataset described in the following. Specifically, we want to derive a proxy for the \emph{structure} of this weighted, and time dependent, network. Further, we want to estimate correlations between OTC derivatives, i.e., infer on possible counterparties from the co-movement of the engagement of institutions.

\section*{The Network of OTC Derivatives}
\label{sec:derivat}

\subsection*{Activities and Ranks}
\label{sec:analysis}

In order to reconstruct the network of counterparty risk from the available dataset, we need to introduce a few variables that are later to be mapped to specific data. 

\begin{figure}[htbp] 
  \begin{center}
    \includegraphics[width=0.95\linewidth]{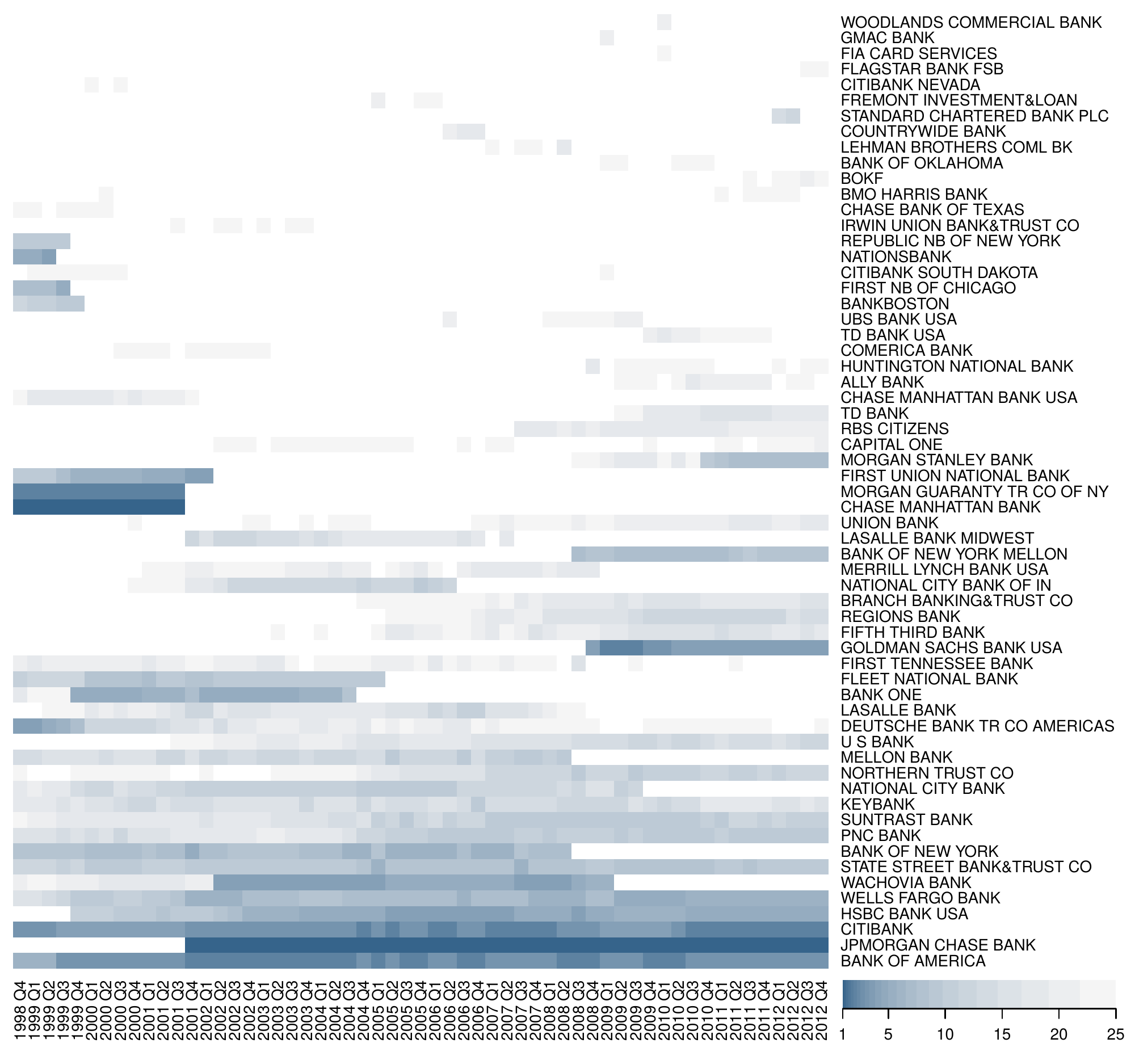}
  \end{center}
  \caption{Time series of the financial institutions appearing among the 25 top ranked between 1998 and 2012. Color codes the rank: the darker the color the better the rank (rank 1 considered the highest), white indicates the absence in the ranking.  
}
  \label{fig:present}
\end{figure}

First of all, we identify each institution in the dataset by an index $i=1,...,N$, where $N=61$, i.e., the total number of distinct institutions. Note that the dataset for each quarter only lists the 25 best ranked institutions, which are not necessarily the same for each quarter (see also Fig.~\ref{fig:present}). 
Thus, during the whole period of 14 years, 61 different institutions appeared in the dataset. 

At each time step $t$, where $t$ is discrete and measured in quarters, up to $T=57$, institutions $i$ and $j$ can act as counterparties, i.e., they have contracts of total volume $x_{ij}(t)$. Importantly, the dataset neither lists the counterparties $j$ nor the volume of their contracts, $x_{ij}(t)$. It lists, however, the quarterly activity of each institution, $a_{i}(t)=\sum_{j=1}^{N}x_{ij}(t)$, i.e., the \emph{aggregated volume}, given in column 5 of Table~A in \nameref{S1_Appendix} Thus, the aim of our paper is to reconstruct the network of dependencies from this aggregated data. Note that, if an institution was not active in a particular quarter, i.e., not listed in the dataset for that period, its activity is set to zero. 

\begin{figure}[htbp]
  \begin{center}
     \includegraphics[width=.55\linewidth]{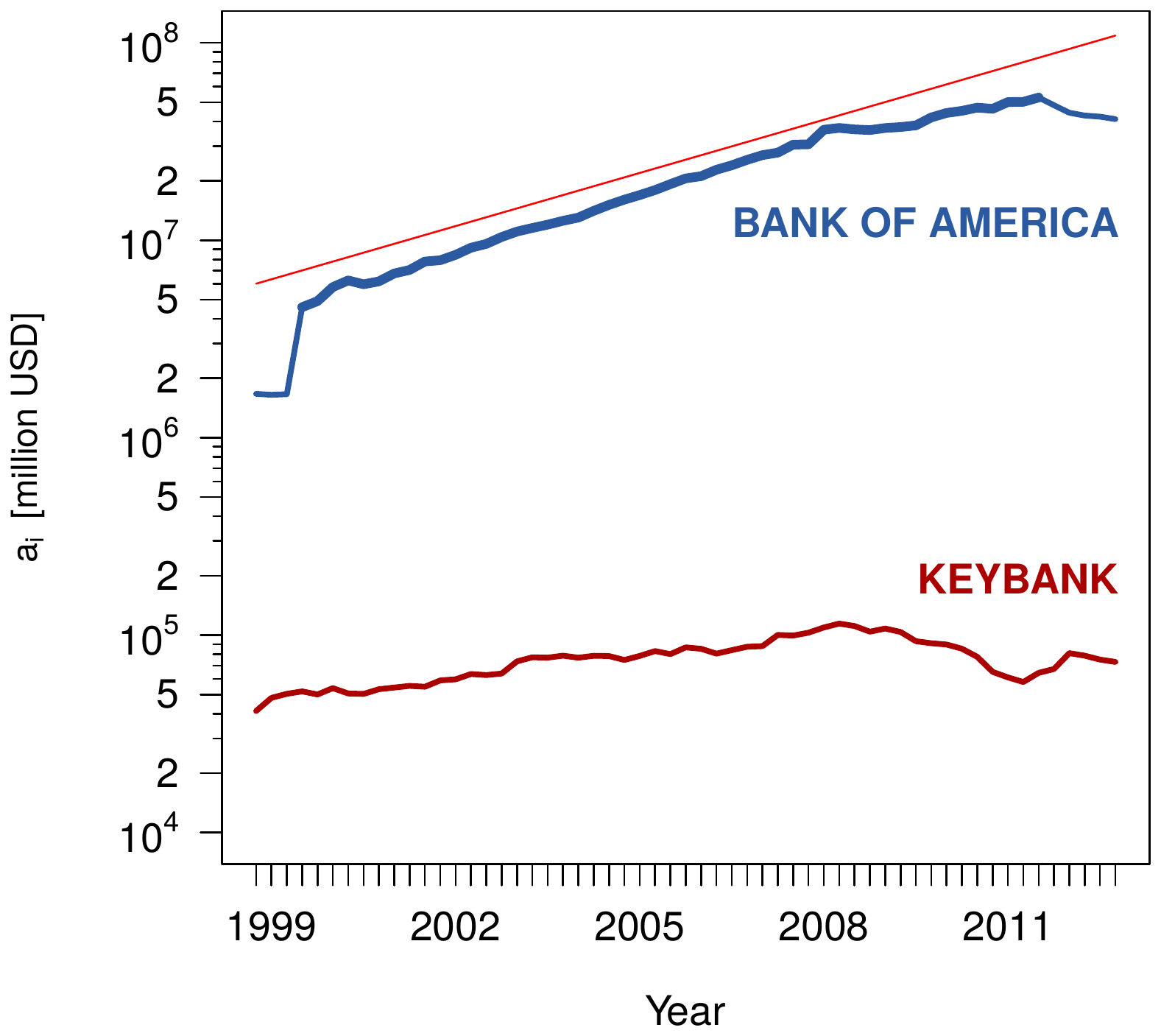}
   \end{center}
   \caption{The total derivatives notional amount of two banks which constantly appear during the whole period from 1998 to 2012. The difference of order of magnitude motivates to take into account the ranks of institutions when building their network. 
The linear regression slope for $log(a_{BoA})$ for the  period 1999/Q3 -- 2011/Q3 (bolder line) is $0.206638$, which corresponds to yearly growth ratio ($a_t(t+1)/a_i(t)$) equal to $1.229537$.
}
  \label{fig:corrcomp}
\end{figure}

To give an example, Fig.~\ref{fig:corrcomp} shows the activity of two banks that are consistently engaged in OTC derivatives in every quarter. Impressively enough, their activities differ in about \emph{two orders of magnitude} and further show a different business strategy over time. While the quarterly activity of \textit{Keybank} remains almost constant over 12 years, the activity of \textit{Bank of America} grew \emph{exponentially} during the same period of time, clearly shown in the linear slope in the logarithmic plot. Only in 2012, after the financial crisis, this involvement was slightly reduced.   

Based on the quarterly activities, $a_{i}(t)$, we can assign each institution $i$ a rank $r_{i}(t)\leftarrow r[a_{i}(t)]$ with $r$ discrete and $r\in \{1,2,...N\}$ such that $r[a_{i}(t)] < r[a_{j}(t)]$ if $a_{i}(t) > a_{j}(t)$ for any pair $i,j \in N$. I.e., rank 1 corresponds to the institution with the highest activity value at time $t$, rank 2 to the one with the second highest activity, and so forth. If an institution was not active in a given period, its rank is set to zero. 
Because the rank $r_{i}$ considers the position \emph{relative} to other institutions, it can change even if the activity of an institution remained constant over a certain period. 

\begin{figure}[htbp]
  \centering
  \includegraphics[width=0.855\textwidth]{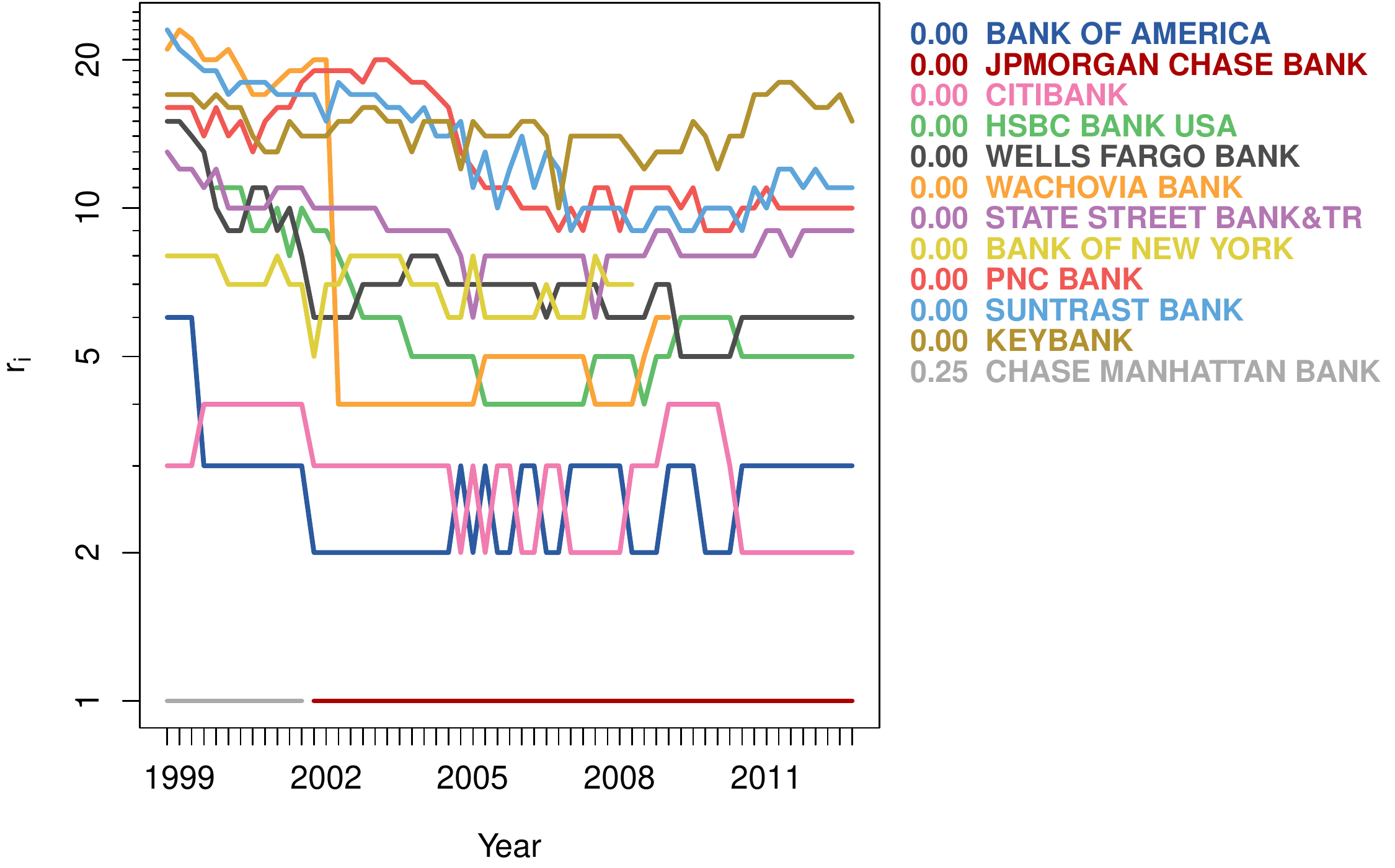}
  \caption{Changes of the ranks $r_{i}(t)$ of a set of banks, with the number showing their distance to the core of the weighted network based on the co-occurrence and activity of financial institutions introduced in \nameref{sec:link}.}
  \label{fig:rank-t}
\end{figure}

Fig.~\ref{fig:present} gives an overview of how often the institutions were present in the ranking up to 25 in any of the quarters, with their ranks color coded. This matrix already indicates that there are remarkable fluctuations in the ranks of most of the institutions, except for a group of about 10 institutions. Fig.~\ref{fig:rank-t} gives a more detailed picture by plotting the ranks of this group over time. We observe that there exists a smaller core group (of about 7 members) with consistently low ranks, which can be well separated from a second group with higher, and more fluctuating, ranks. 

\begin{figure}[htbp]
  \centering
  \includegraphics[width=0.95\textwidth]{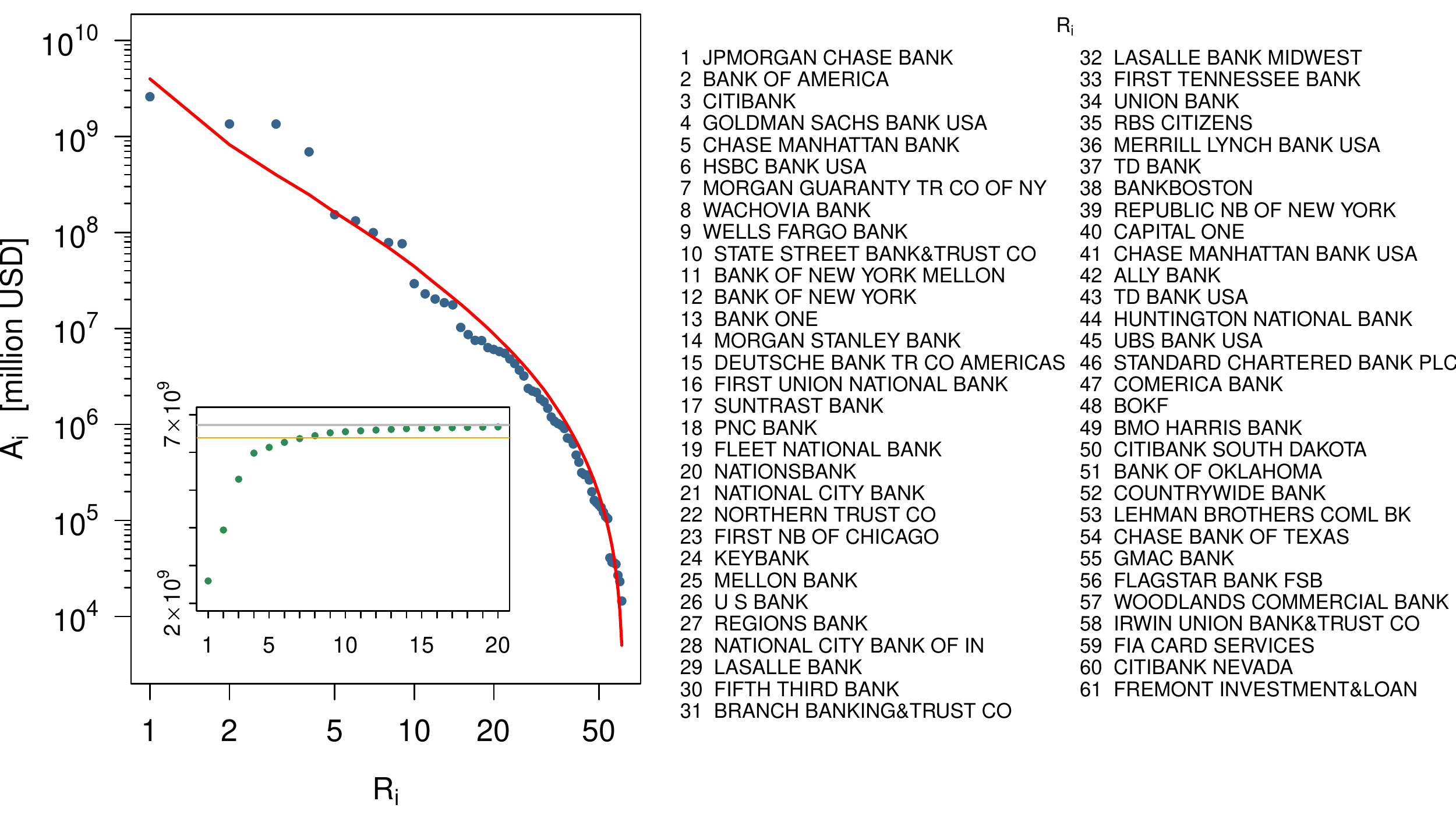}
  \caption{Distribution of the aggregated activity  $A_{i}$ over the rank $R_{i}$ obtained from the whole reporting period. (inset)  Cumulative sum  $P(R<Y)=\sum_{R=1}^{Y}A(R)$. The ceiling of the distribution, which is the capacity of the market over the whole period of time is shown by the grey line, while the orange line shows the corresponding 95\% percentile. 
}
  \label{fig:agg-rank}
\end{figure}

This can be also observed by looking at the ranks $R_{i} \leftarrow r[A_{i}]$ resulting from the aggregated activities $A_{i}=\sum_{t=1}^{T}a_{i}(t)$. Plotting the inverse function $A(R)$ shown in Fig.~\ref{fig:agg-rank}, we observe a rather skew distribution of the aggregated activities with respect to the rank, with a skewness value $\gamma=4.637150$ and a Gini coefficient \cite{Gini1909} $g=0.9558996$. Moreover, the plot suggests that the aggregated activity $A$ follows a log-normal distribution with respect to the rank $R$:
\begin{equation}
  \label{eq:log-normal}
  A(R) = \frac{1}{R \sigma \sqrt{2 \pi}} \cdot \exp\left[- \frac{\left(\ln R
        - \mu\right)^2}{2 \sigma^{2}}\right]\;;\;\;
      R \geq 1
\end{equation}
where $\mu=14.54116$ is the mean value and $\sigma=2.865165$ the standard deviation of the distribution. 
To further compare the empirical with the log-normal distribution, Fig.~A in \nameref{S1_Appendix} shows the $Q-Q$ plot and gives the results of the two-sample Kolmogorov-Smirnov test.  

The inset of  Fig.~\ref{fig:agg-rank} presents the cumulative distribution $P(R<Y)=\sum_{R=1}^{Y}A(R)$. It indicates that 
 about 95\% of the total activity results from  the seven first ranked institutions, while the 15 first ranked institutions cover more than 99\% of the total activity. It may be tempting to restrict the analysis to only these 15 institutions. However, the aggregated activities do not allow to draw conclusions about the concentration of activities in certain time periods or a change of strategy in choosing counterparties, before and after the financial crisis. Therefore, we will present more details on the temporal activities in Section \nameref{sec:link}.

The available data also allows us to analyse the composition of the activities $a_{i}(t)$ with respect to exchange traded derivatives (ETD) and OTC derivatives. 
I.e., the value of the total derivatives  
is split into $a_{i}(t)=a_{i}^{\mathrm{ETD}}(t)+a_{i}^{\mathrm{OTC}}(t)$ and $A_{i}=A_{i}^{\mathrm{ETD}}+A_{i}^{\mathrm{OTC}}$, respectively. Already the sheer numbers of the $a_{i}(t)$ and $a_{i}^{\mathrm{OTC}}(t)$ tell that OTC derivatives make up for the vast amount of contracts. I.e., we should not assume that the ranks $r_{i}(t)$ or $R_{i}$ obtained from both ETD and OTC derivatives are different from those ranks that would result from only considering the values of $a_{i}^{\mathrm{OTC}}(t)$ or $A_{i}^{\mathrm{OTC}}$. To test this hypothesis, Fig.~B
in the \nameref{S1_Appendix} provides a $Q-Q$ plot to compare both values. We see that up to rank 15 there is no difference in the ranks obtained by these two measures, whereas between ranks 15 and 50 the difference in ranks would be 1 or 2. Only for ranks above 50, the differences become remarkable. So it is reasonable to use the ranks $r_{i}(t)$ and $R_{i}$ in the further evaluation. 

\begin{figure}[htbp]
  \centering
  \includegraphics[width=0.5\textwidth]{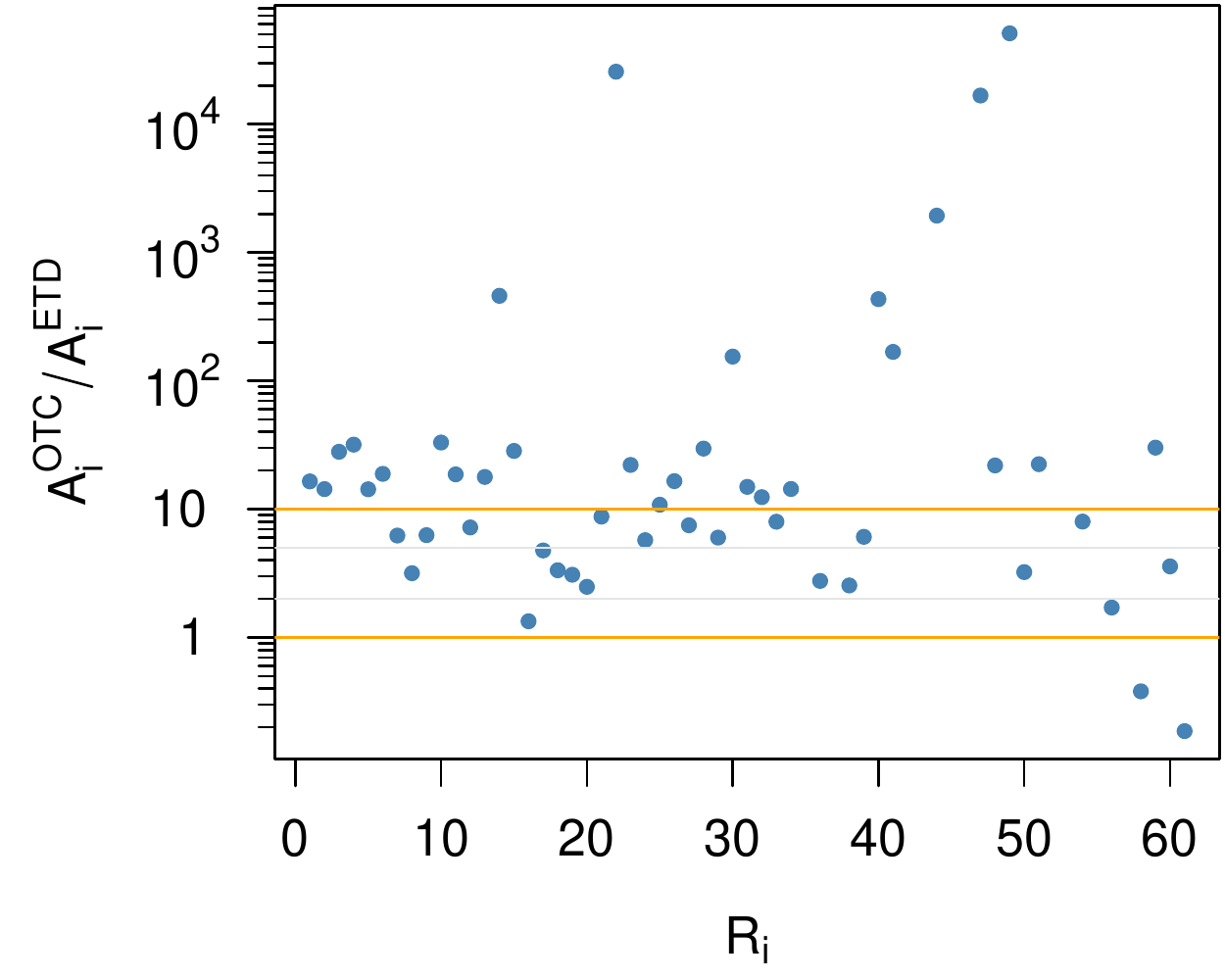}
  \caption{Ratio $A_{i}^{\mathrm{OTC}}/A_{i}^{\mathrm{ETD}}$ versus ranks $R_{i}$ based on the total activity $A_{i}$.
}
  \label{fig:otcratio}
\end{figure}

However, when analysing the counterparty risk in derivative contracts, we will make a distinction between the (less risky) ETD and the more risky OTC derivatives. In fact, as Fig.~\ref{fig:otcratio} indicates, the importance of OTC derivatives as compared to the ETD vastly differs across institutions. The ratio 
$A_{i}^{\mathrm{OTC}}/A_{i}^{\mathrm{ETD}}$ is below 10 for about 1/3 of all institutions, which implies that 10\% or more of the activities is in ETD. However, looking at the 15 best ranked institutions, we see for most of them the ETD business accounts for only 2\%-5\% of their activity. So again, it is reasonable to proxy activities related to  OTC derivatives by the total activities - but whenever possible, we will take into account the real values for OTC derivatives.

\subsection*{Temporal and aggregated networks}
\label{sec:link}

In order to estimate the link structure of the network of counterparty risk, we first look into the co-occurrence of any two institutions among the 25 best ranked institutions in each given quarter. I.e., we define a link as $l_{ij}(t)=1$ if for both institutions $1\leq \{r_{i}(t),r_{j}(t)\}\leq 25$ and $l_{ij}(t)=0$, otherwise. Their co-occurrence does not necessarily imply that the two institutions are counterparties of an OTC derivative. A ranked institution $i$ could do all its OTC contracts with the many institutions that have ranks too high (i.e., activities too low), to be listed in this dataset. Practically, however, this cannot be the case because, as the OCC reports verify, already 99\% of all OTC derivatives are held by the 25 best ranked institutions. So, the not listed ones would make only for 1\%, which cannot explain the large activities of any of the 25 best ranked institutions. Consequently, it is reasonable to assume that $i$ has at least one contract with any of the other 24 institutions, and the best ranked institutions have likely more than one. 

The co-occurrence network certainly overestimates the business relations based on OTC contracts because it is basically a fully connected network between the 25 best ranked institutions. Further, the co-occurrence may change in each quarter. Therefore, as the next step, it is reasonable to assign weights for the links between any two institutions based on the number of quarters, they co-appear in the dataset. I.e., we define weights as
\begin{equation}
w_{ij}= \frac{1}{T} \sum_{t=1}^{T}l_{ij}(t)
\label{eq:wij}
\end{equation}
to normalize them to the available time period. A node that has links with high weights to its neighbors certainly represents an important institution in the OTC derivatives market. We use the weights to define the importance of an institution as $W_{i}=\sum_{j=1}^{N}w_{ij}$. In the following network figures, the size of the nodes is scaled to the \emph{normalized} importance, $W_{i} /\sum_{i} W_{i}$.   

\begin{figure}[htbp]
  \begin{center}
    \includegraphics[width=0.85\linewidth]{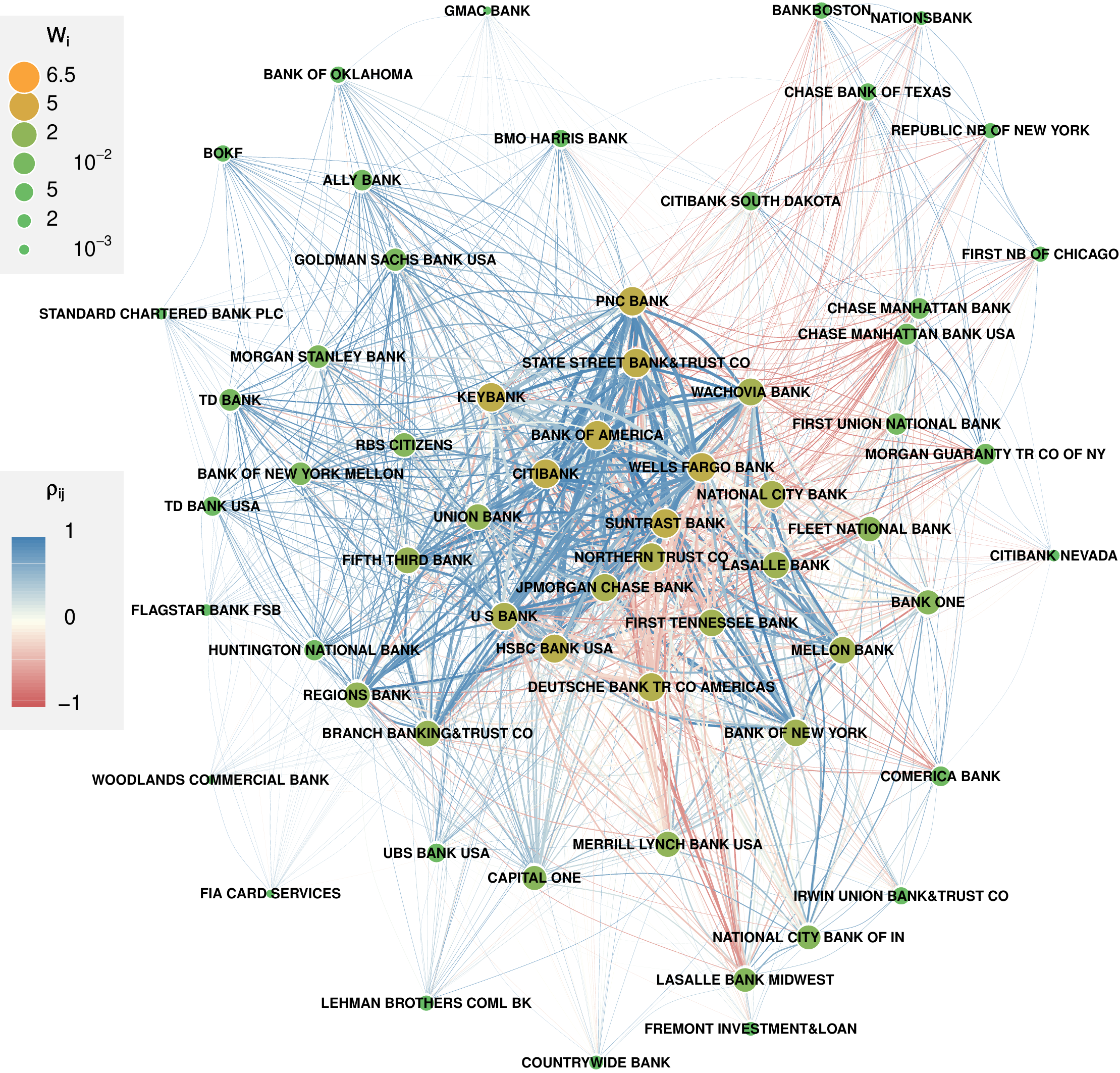}
  \end{center}
  \caption{Weighted network based on the co-occurrence of financial institutions in the top 25 ranking, aggregated over all quarter years. The size of a node increases with its importance $W_{i}$, the width of the links increases with their weights $w_{ij}$, where $l_{ij}\in \{0,1\}$ (i.e., do not depend on the ranks).
The links are colored according to the non-normalized correlation coefficient (defined in Section \nameref{sec:corr-act}) between activities in OTC derivatives of the two banks.
}
  \label{fig:cooccurrence}
\end{figure}

This allows us now,  based on the aggregated values, to draw in Fig.~\ref{fig:cooccurrence} a first approximation of the network of counterparty risk. 
While this figure clearly shows the important institutions with respect to their \emph{co-occurrence}, it neglects another important information, namely their \emph{ranking} which is a proxy of their relative  \emph{activity}.
Imagine institution $i$ with a steady but relatively low  activity over time, just enough for frequently appearing in the network, while institution $j$ may have a much higher activity, but during a shorter period of time, resulting in a better, but less frequent ranking. As a result, institution $i$ will be over-presented in the network drawn in Fig.~\ref{fig:cooccurrence}, while institution $j$ will be under-represented. Such activity differences are prevalent in the dataset as the investigations in Section \nameref{sec:analysis} show. In the example shown in Fig.~\ref{fig:corrcomp}, the activity of \textit{Keybank} was two to three orders of magnitude lower than the activity of  \textit{Bank of America}. But  because \textit{KeyBank} was present in the top 25 list during the whole time period, it gained a similar position in the network in Fig.~\ref{fig:cooccurrence} as giants such as \textit{Bank of America} or \textit{Citibank}. 

Therefore, to further improve our estimation of the network of counterparty risk, we take into account the overall activity of an institution by using their ranks to assign weights to the links of co-occurrence. I.e., instead of $l_{ij}(t)=1$, we use
\begin{equation} 
\label{eq:lijweight}
  l_{ij}(t)=\min{\left\{\frac{1}{r_{i}(t)},\frac{1}{r_{j}(t)}\right\}}
\end{equation}
The rationale behind is to bind the weight of a link to the activity of the \emph{less active} institution. To elucidate this, let us assume that institution $i$ is a big player with rank $r_{i}(t)=2$ at time $t$, while $j$ is a less important institution  with rank $r_{j}(t)=21$. Because both institutions co-appear in the same quarter, each of them has links to all other institutions listed in the same time period, i.e., 24 links. For the less important institution $j$, 20 of these links get assigned a weight of $1/20$, namely those links to institutions with better ranks. But there are 4 links to institutions with an activity less than $j$ and therefore with higher ranks. Those links get assigned the weights $1/22$, $1/23$, $1/24$, $1/25$. I.e., for each institution, links to less active counterparties have less weight, while links to more active counterparties have the maximum weight that could occur given the rank of that institution. Likewise, for institution $i$ only one link, namely the link to the highest ranked institution, gets a weight $1/2$, whereas the 23 links to all other institutions become less and less important as $1/3$, $1/4$, ..., $1/25$. 

The resulting network is shown as an animation (at the time of writing only supported in Adobe\tsp{\textregistered} products) in Fig.~D in \nameref{S1_Appendix}. At each time step this is a fully connected network, but the weights of the links, as well as the importance of the institutions, change during every timestep. 
The animation nicely elucidates the emergence of new key players in the OTC derivatives markets before and after the crisis, as well as the changed preferences in choosing counterparties.

\begin{figure}[htbp]
  \begin{center}
    \includegraphics[width=0.85\linewidth]{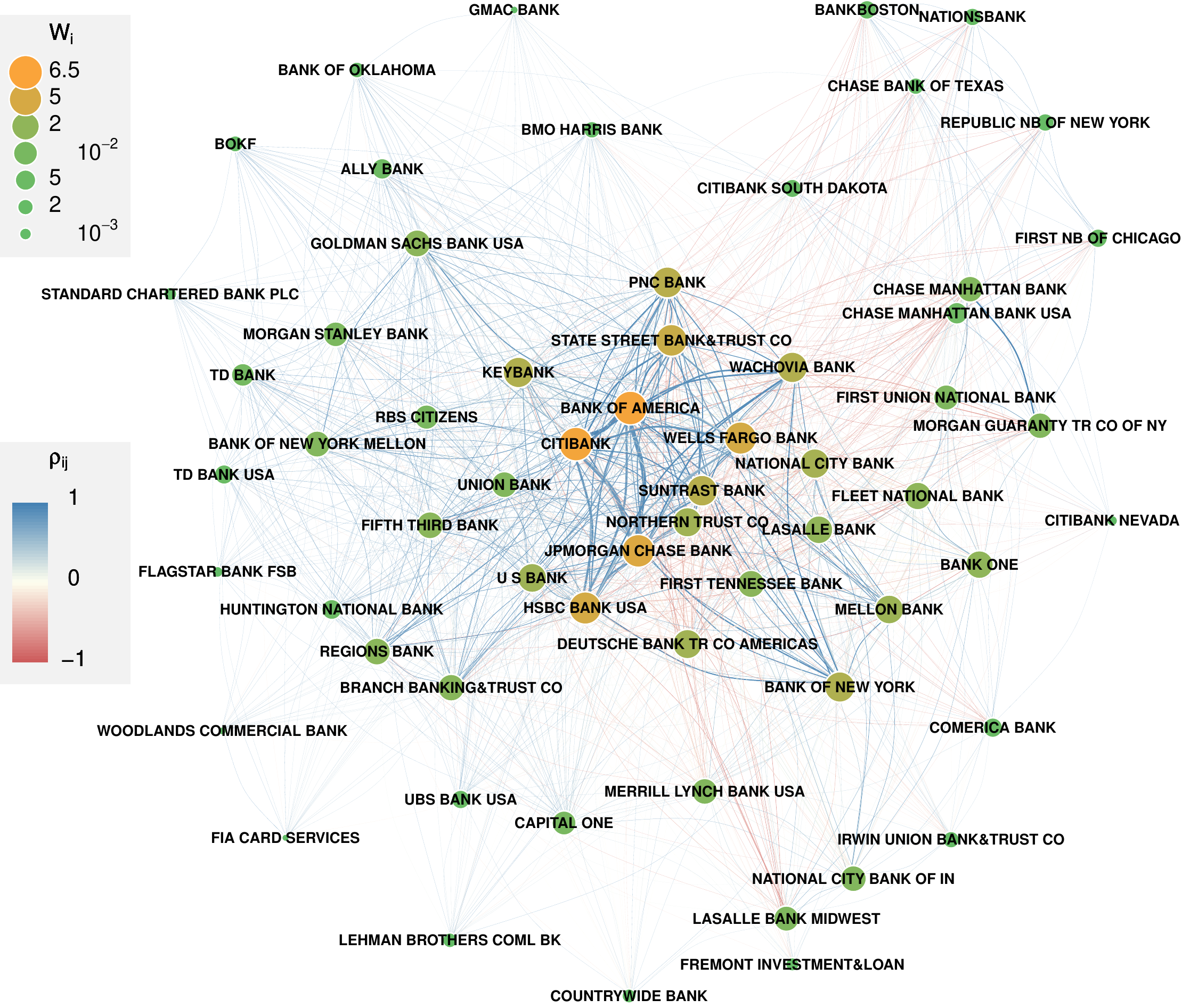}
  \end{center}
  \caption{
Weighted network based on the co-occurrence and activity of financial institutions in the top 25 ranking, aggregated over all quarter years. The coding of size and color of nodes and links are the same as in Fig.~\ref{fig:cooccurrence}, but the $w_{ij}$ and  $W_{i}$ are calculated from the $l_{ij}$ as given by Eq.~(\ref{eq:lijweight}) i.e., dependent on the ranks. The time resolved network is shown in Fig.~D in \nameref{S1_Appendix}. The aggregated network should be compared with Fig.~\ref{fig:cooccurrence} where activities are not taken into account. 
}
  \label{fig:weightednet}
\end{figure}

To allow a comparison with Fig.~\ref{fig:cooccurrence}, we aggregate the weights of the links over time according to Eq.~(\ref{eq:wij}), to take into account both  co-occurrence and activity, and  calculate the importance of an institution as before,  $W_{i}=\sum_{j=1}^{N}w_{ij}$. The resulting weighted network is then shown in Fig.~\ref{fig:weightednet}, which should be compared to  Fig.~\ref{fig:cooccurrence}. 
The most obvious difference is a less dense core, built up by a smaller number of  important institutions,  in Fig.~\ref{fig:weightednet}.  Tracing particular institutions, e.g. 
\textit{Union Bank}, 
we see that their position becomes less influential.    
 But the core of the network, i.e., the set of the ten most important institutions, remains the same and shall be investigated in the following.

\subsection*{Core-periphery structure}
\label{analysis:1}

So far, we have used the following information to describe counterparty relations: (i) \emph{Aggregated measures} derived from the aggregated \emph{co-occurrence} $l_{ij}$ in the ranking of the 25 top players in the OTC market, in particular the \emph{weights} $w_{ij}$ and the \emph{importance} $W_{i}$. The results are concluded in the network of Fig.~\ref{fig:cooccurrence}. (ii) \emph{Temporal measures} derived from the \emph{ranking} $r_{i}(t)$, in particular the temporal \emph{co-occurrence} $l_{ij}(t)$. The results are concluded in the animated network of Fig.~D in \nameref{S1_Appendix}, with the time-aggregated network shown in Fig.~\ref{fig:weightednet}. While the latter  can be seen as the most refined network of counterparty risk, the characterization of both nodes and links is  still  based on the \emph{activity} $a_{i}(t)$  of the respective institution, i.e., it is derived from a single scalar measure. So, the question is whether the reconstruction of the aggregated temporal network would allow us to add another dimension to characterize institutions, based on \emph{topological} information. 

\begin{figure}[htbp]
  \begin{center}
    \includegraphics[width=0.75\linewidth]{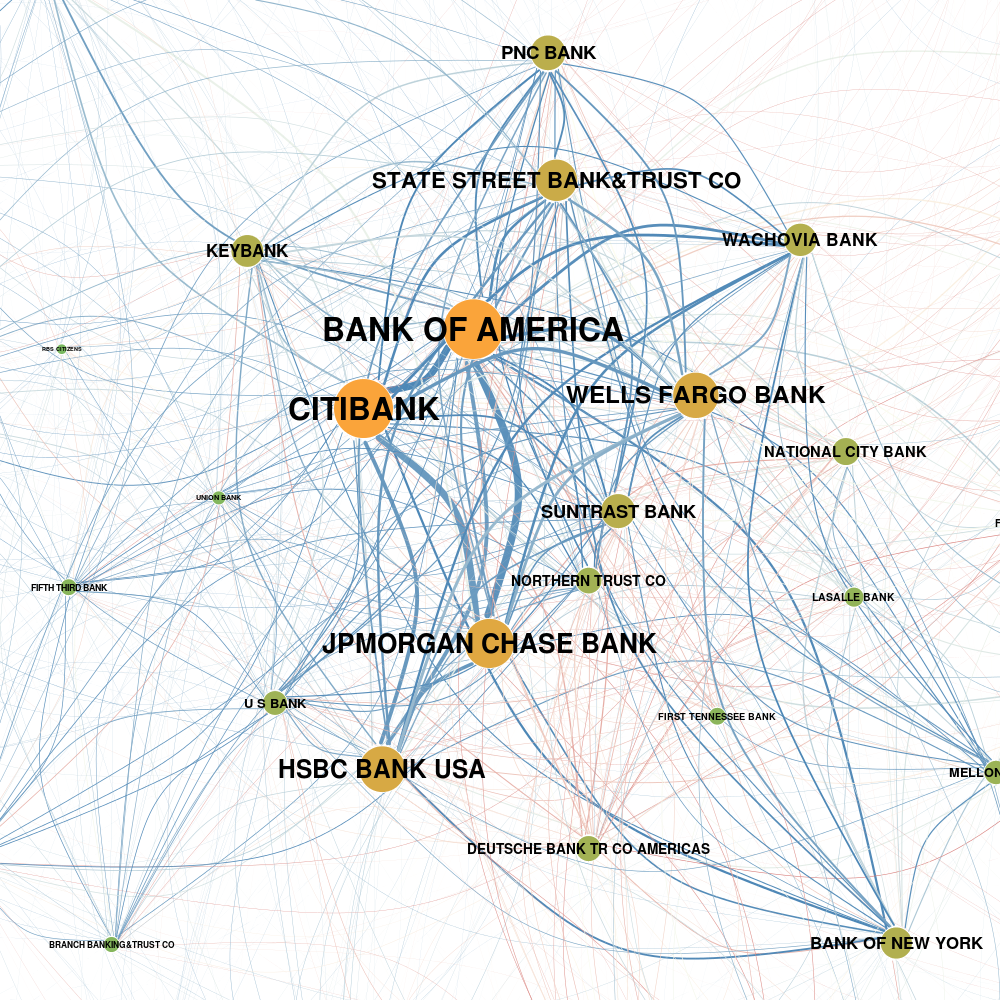}
  \end{center}
  \caption{The core of the aggregated weighted  temporal network presented in Fig.\ref{fig:weightednet}. For the coding see the legend in Fig.\ref{fig:weightednet}.}
  \label{fig:net-core}
\end{figure}

Already a visual inspection of Figs. \ref{fig:cooccurrence} and \ref{fig:weightednet} verifies that the network is  rather heterogeneous with respect to its density. We can easily detect a \emph{core} of larger (i.e., more active) and more densely connected nodes which can be distinguished  from a \emph{periphery} of nodes that are smaller (i.e., less active) and less densely connected. In fact, peripheral nodes are mostly  connected towards the core and much less to other peripheral nodes. 
The core of the network is depicted in Fig.~\ref{fig:net-core} and gives a good impression of the fully connected network, albeit with links of different weights. 

Whether institutions can be found in the core or in the periphery of the network certainly relates to their importance in the OTC market. In order to quantify the topological information encoded in the network structure, we use the \emph{weighted $K$ core analysis}, which is an established method to assign an importance value to nodes. In the first step, for the time aggregated network shown in Fig.~\ref{fig:weightednet}, each node gets assigned a \emph{weighted degree} $\hat{k}_{i}$~\cite{Garas2012a}: 
\begin{equation}
  \hat{k}_{i}=\left[k_{i}^{\alpha} \left( \sum_{j}^{k_{i}}{w_{ij}}
    \right)^{\beta} \right]^{\frac{1}{\alpha+\beta}}, \nonumber
  \label{eq:gsh.kprime}
\end{equation}
where $k_{i}$ is the degree of node $i$, i.e., its number of links to neighboring nodes, and $\sum_{j}^{k_{i}}{w_{ij}}$
is the sum over all its link weights as defined in Eqn. (\ref{eq:wij}) with the weighted $l_{ij}$ given by Eqn. (\ref{eq:lijweight}). The exponents $\alpha$ and $\beta$ are used to weight the two different contributions, i.e., \emph{number} of links versus \emph{weight} of links. 
In our analysis we used $\alpha = 0$ and $\beta =1$, i.e., we focused only on the weights since the network is almost fully connected and the node degree does not give us any information.

In the second step, we follow a pruning procedure to recursively remove all nodes with degree $\hat{k} \leq K$ from the network, where $K=1,2,....$. I.e.,  first all nodes with  $\hat{k} \leq 1$ are removed, which may leave the network with other nodes that now have  $\hat{k} \leq 1$ simply because some of their neighbors were removed. So the procedure continues with removing these nodes, too, unless \emph{no} nodes with $\hat{k} \leq 1$ are left. 
Then all nodes removed during this step get assigned to a \emph{core} $K=1$, and the procedure continues to successively remove all nodes with degree $\hat{k}\leq 2$ and assign them to a core $K=2$, etc. The procedure stops at a certain high core value, $K$, when all nodes are removed. The higher the $K$-core a node is assigned to, the more it belongs to the ``core'' of the network and the more important it is, from a topological perspective.  Evidently, nodes assigned to a core with low $K$ value are much less \emph{integrated} in the network. This does not refer simply to the number of neighbors, but also to non-local properties such as the number of neighbors of their neighbors, because the $K$-core decomposition also takes these into account. That means, the $K$-core a node is assigned to reflects is position in the network much better than simple measures such as the degree (i.e., the number of neighbors), alone. 

The results of the weighted $K$-core analysis are shown in the left side of Fig.~C in \nameref{S1_Appendix}, where the $K$ value is normalized to $1$. Based on their $K$ value, institutions can be ranked such that the higher the $K$ value (i.e., the better the integration in the network), the better the rank. This \emph{topological}  ranking does not necessarily coincides with the ranking $R_{i}$ obtained from the aggregated \emph{activity} $A_{i}$ which is shown on the right side of Fig.~C in \nameref{S1_Appendix}, for comparison. 
This indicates that \emph{structural} measures based on the network \emph{topology} indeed provide information different from the \emph{temporal} measures based on the market activities of the institutions. But, comparing the left and the right sides with respect to the color coding, we observe that only in a few cases institutions have considerably different levels of importance dependent on the measurement. It would be worth looking at these in a case-by-case study, to find out which importance measure better reflects their overall performance in the financial market.

We note that, for consistency, we have used the ranking obtained from the weighted $K$-core analysis to sort the different institutions in the figures, in particular in Fig.~\ref{fig:present} where their presence in the dataset is given, and in Figs. \ref{fig:corrfin} and \ref{fig:correxposure} where the correlations in activities and in risk are presented.

\section*{Correlations}
\label{sec:correlations}

\subsection*{Correlation measures}
\label{sec:corr-measures}

So far, we have analysed the \emph{co-occurrence} of financial institutions in the set of the 25 best ranked institutions, weighted by their ranks. These ranks were based on their activities, i.e., \emph{total derivatives}. As a result, we could reconstruct the weighted network of counterparty risk which also reflects the importance of the nodes. This network was reconstructed (a) on a time resolution of one quarter year, to show the dynamics of the network (Fig.~D in \nameref{S1_Appendix}), and (b) on the time aggregated level (Fig.~\ref{fig:weightednet}).  

To further analyse the mutual dependence between the best ranked institutions, we now calculate different correlations. The network of counterparty risk has revealed how the co-occurrence changes over time. But will the OTC derivatives of institution $i$ increase, or decrease, if the same measure of institution $j$ increases? Answering this question allows some more refined conclusions about the dependence between these institutions. 

The simplest measure is the \emph{Pearson correlation coefficient} $\rho$, which points to a \emph{linear} dependence between two variables. As explained above, for each institution $i$ we have a dataset $\mathbf{a}_{i}=\{a_{i}(1),a_{i}(2),...,a_{i}(T)\}$ available which contains up to $T$ entries about its quarterly activity $a_{i}(t)$ measured by means of its total derivatives. We recall that some of these entries are zero whenever institution $i$ was not listed among the best 25 ranked. Let us define the mean value and the standard deviation of each of these samples as: 
    \begin{equation}
      \label{eq:mean}
  \bar{a}_{i}=\frac{1}{T}\sum_{t=1}^{T} a_{i}(t)\;; \quad 
s^{a}_{i}=\sqrt{\frac{1}{T-1}\sum_{t=1}^{T} \big[a_{i}(t)-\bar{a}_{i}\big]^{2}}.
    \end{equation}
The Pearson correlation coefficient with respect to the variable $a$ is then defined as
\begin{equation}
  \label{eq:pearson}
  \rho^{a}_{ij} = \frac{1}{T-1} \sum_{t=1}^{T} \left[\frac{a_{i}(t)-\bar{a}_{i}}{s^{a}_{i}}\right]
 \left[\frac{a_{j}(t)-\bar{a}_{j}}{s^{a}_{j}}\right].
\end{equation}
Values of $\rho$ can be between -1 and +1. The latter indicates that the relation between activities $a_{i}$ and $a_{j}$ can be perfectly described by a linear relationship, where $a_{i}$ increases as $a_{j}$ increases. -1, on the other hand, indicates a perfect linear relationship where $a_{i}$ decreases as $a_{j}$ increases, and vice versa. Zero would indicate that there are no linear dependencies detected in the data. Eq.~(\ref{eq:pearson}) also shows that, in case of a positive correlation, if $a_{i}(k)>\bar{a}_{i}$ then also $a_{j}(k)>\bar{a}_{j}$ for most of the time, and if $a_{i}(k)<\bar{a}_{i}$ then also $a_{j}(k)<\bar{a}_{j}$ for most of the time, i.e., the activities of both institutions are mostly above (or below) their respective average, at the same time.

\subsection*{Correlations in activities}
\label{sec:corr-act}

\begin{figure}[htbp]
  \begin{center}
    \includegraphics[width=.56\linewidth]{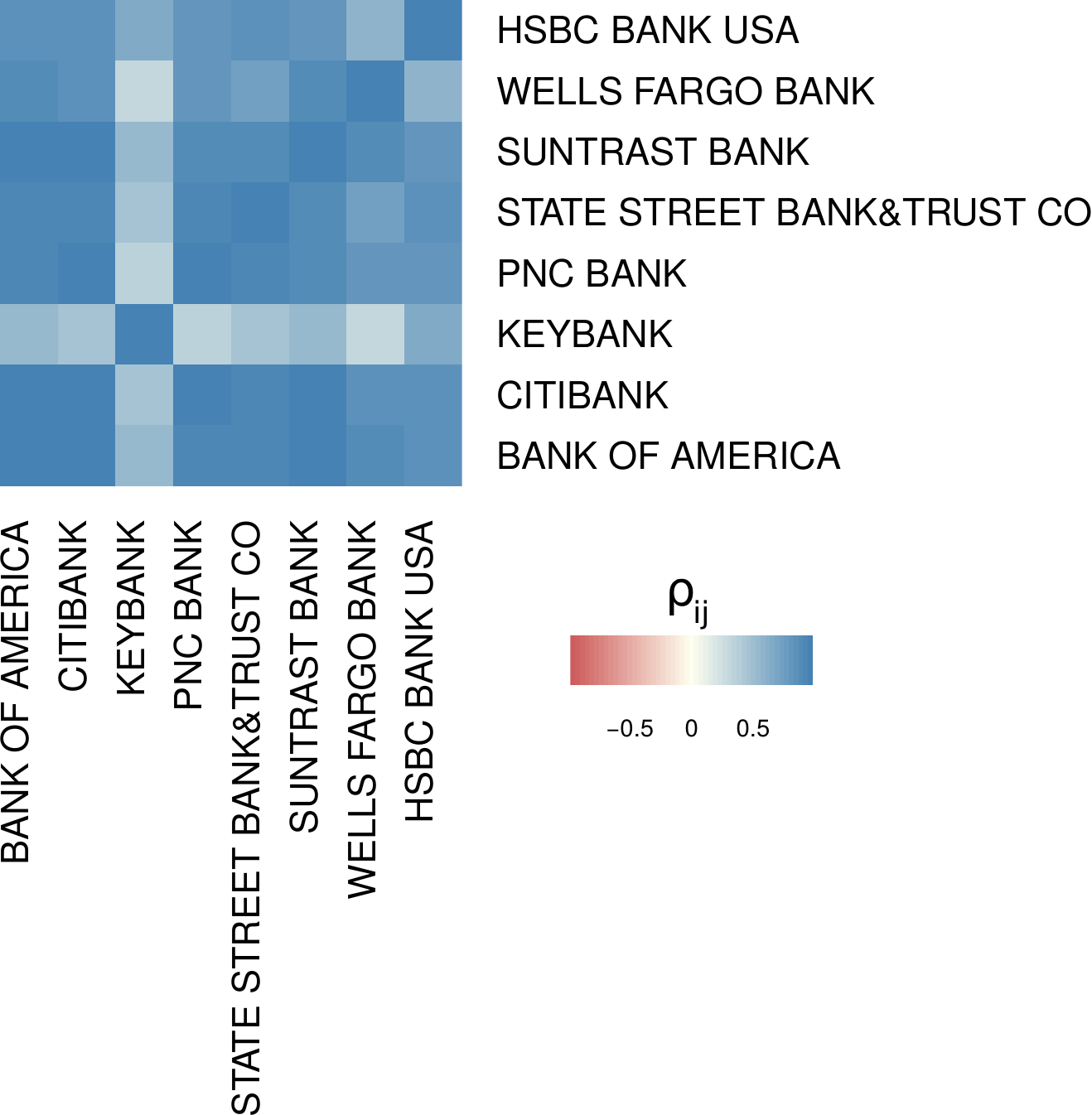}
  \end{center}
  \caption{Correlation matrix of the reported total derivatives of the institutions appearing in top 25 commercial banks, savings associations or trust companies in derivatives during the whole period from 1998 to 2012.}
  \label{fig:corrsimple}
\end{figure}

We first discuss the results for the most active institutions, i.e., those appearing  among the 25 best ranked institutions with respect to their total derivatives in every quarter. Interestingly, this applies only to 8 out of the 61 listed institutions. 
Fig.~\ref{fig:corrsimple} shows the correlation matrix for these institutions,  their activities proxied by the total notional amount of derivative contracts as listed in column 5 of Table~A in \nameref{S1_Appendix}.

There are two observations to be made: (i) the correlations between any two of these institutions are always positive and often even close to 1, (ii) \textit{Keybank} is a noticeable exception. This can be explained by the combination of two effects: The first one is the vastly growing market in OTC derivative during the observation period which resulted in the growth of OTC derivatives for these core institutions. Thus, the observed correlations could, in principle,  be caused by the underlying market dynamics rather than by the mutual interaction. However, taking into account that the 10 best ranked institutions already account for 95\% of the OTC derivatives market, there is little room for the assumption that their growth is based on OTC derivative contracts with institutions that do not belong to the core of 10, or to the 25 best ranked institutions. In conclusion, these eight institutions increased their OTC derivatives activities by repeatedly choosing the same core institutions as counterparties. The low correlations for \textit{Keybank} could result both from the absence of growth (see Fig.~\ref{fig:corrcomp}), while all others were growing, and from choosing counterparties from outside the set of core banks. 

If we wish to extend this correlation analysis to the whole set of 61 institutions, it would generate a number of artifacts which should be avoided. We discuss them here, first, to motivate our own approach presented afterward. 
As already shown in Fig.~\ref{fig:present}, most of these institutions were not present in the ranking of the best 25, for some longer or shorter period. So, one could limit the correlation analysis to those quarters where the two institutions were indeed present in the ranking.  I.e., if institution $i$ appeared at times $t_1, t_2, t_3, t_4$ and while institution $j$ appeared at times $t_2, t_4, t_5, t_6$, the correlation coefficient for them is computed using only the observations at times $t_2$ and $t_4$ where both were present. The Pearson correlation coefficient based on pairwise available observations with respect to the variable $a$ is then defined as
\begin{equation}
  \label{eq:pearsonT}
  \rho^{a}_{ij} = \frac{1}{\#[\mathcal{T}_i \cap \mathcal{T}_j]-1} \sum_{t\in \mathcal{T}_i \cap \mathcal{T}_j} \left[\frac{a_{i}(t)-\bar{a}_{i}}{s^{a}_{i}}\right]
 \left[\frac{a_{j}(t)-\bar{a}_{j}}{s^{a}_{j}}\right]
\end{equation}
$\mathcal{T}_i$ and $\mathcal{T}_{j}$ are subsets of $\{1,2,...,T\}$, comprising the time steps when the institutions $i$ and $j$ appeared in the ranking among the top 25, and $\#[\mathcal{T}_i]$ and $\#[\mathcal{T}_i]$ are the numbers of these time steps.
$\mathcal{T}_i \cap \mathcal{T}_j$ then defines the subset of timesteps where \emph{both} institutions $i$ and $j$ appeared together, and $\#[\mathcal{T}_i \cap \mathcal{T}_j]$ gives the respective number of those time steps. 
Consequently, the average activity $\bar{a}_{i}$ and the standard deviation $s^{a}_{i}$ are also calculated only for the subset $\mathcal{T}_i$:
\begin{equation}
  \label{eq:meanT}
  \bar{a}_{i}=\frac{1}{\#[\mathcal{T}_i]}\sum_{t\in\mathcal{T}_i} a_{i}(t)\;; \quad 
  s^{a}_{i}=\sqrt{\frac{1}{\#[\mathcal{T}_i]-1}\sum_{t\in\mathcal{T}_i} \big[a_{i}(t)-\bar{a}_{i}\big]^{2}}
\end{equation}
The results of this analysis are shown in Fig.~E in \nameref{S1_Appendix}. We observe that, in addition to the strong correlations in the core of those institutions always present, there are a lot of strongly \emph{anti-correlated} activities (indicated by rich red) among the low ranked institutions which need to be interpreted, both with respect to the correlation and to the magnitude. We start with the latter.   

Defining the Pearson correlation coefficient according to Eqn. (\ref{eq:pearsonT}) has the drawback that the correlation coefficients for different institutions are no longer normalized to the same number of observations, $T$, as in Eq.~(\ref{eq:pearson}) and thus cannot be compared. Precisely, the correlations between  \textit{Bank of America} and \textit{Citibank}, which were both present in the ranking for $T=57$ quarters will get the same weight as the correlations between
\textit{Citibank Nevada} and \textit{Chase Manhattan Bank USA} which were present together only two times.

The second drawback results from the time lapse between the co-appearance. 
While the times $t_{4}$ and $t_{6}$ in the above example may still be relatively close, the interval between $t_{4}$ and $t_{56}$ would be much longer and, because of the unknown intermediate values, interpretations about  the correlated move of both institutions become highly speculative. 

In contrast to the above example, in which the two intermediaries appear only in a few quarters, but yet co-appear twice, some pairs of intermediaries which are important both by means of long term presence and good rankings, never appeared \emph{together}, for example \textit{Goldman Sachs} and \textit{Bank of New York}, and, as a consequence, the Pearson correlation coefficient is not even defined for them, which is yet another drawback.

One could argue that these drawbacks disappear if we simply keep the normalization $T$, as in Eqs.~(\ref{eq:mean})(\ref{eq:pearson}), and instead assign an activity $a_{i}(t)=0$ whenever an institution $i$ is not present in the ranking. While there is no evidence that the activity was indeed zero, the error produced this way is certainly small because of the very  skew distribution of activities shown in Fig.~\ref{fig:agg-rank}, and both the mean and the standard deviation of the activity are not substantially affected. But it becomes a problem when there is indeed no data because the institution does not exist in certain quarters, e.g.\ because of mergers and acquisitions, as in the case of \textit{Chase Manhattan Bank} and \textit{JPMorgan Chase Bank}.

Additionally, by proceeding like this we would generate another artifact, namely generating artificial correlations between those institutions that are often not in the rankings and, in the worst case, never co-appear. It is in fact the absence of data that generates their correlations, artificially. Taking again the example of \textit{Goldman Sachs} and \textit{Bank of New York}, these two institutions would then appear \emph{anti-correlated} while, in fact, no correlation was defined for them. Thus, solving the above mentioned drawbacks this way would generate yet a different one. 

Consequently, we will go with the correlations defined on the pairwise co-appearance, Eqn. (\ref{eq:pearsonT}), but we compensate for the different normalization by multiplying the correlation coefficients $\rho_{ij}^a$ with the weights $w_{ij}$ defined in Eq.~(\ref{eq:wij}) with $l_{ij}=1$, which is the relative number of co-appearances.
This implies that the correlations between two institutions that rarely co-appeared in the ranking are scaled down. Precisely, after this correction, the weights $w_{ij}$ define the bounds of the values of the correlation coefficients, which are different for each pair of institution, namely $[-w_{ij}, +w_{ij}]$ instead of $[-1, +1]$. 
These weighted correlation coefficients shall be interpreted differently from the conventional correlation coefficients in that a close-to-zero coefficient no longer means that the variables are uncorrelated, but that there is no \emph{significant} correlation because of the low weight.

\begin{figure}[htbp]
  \begin{center}
    \includegraphics[width=.75\linewidth]{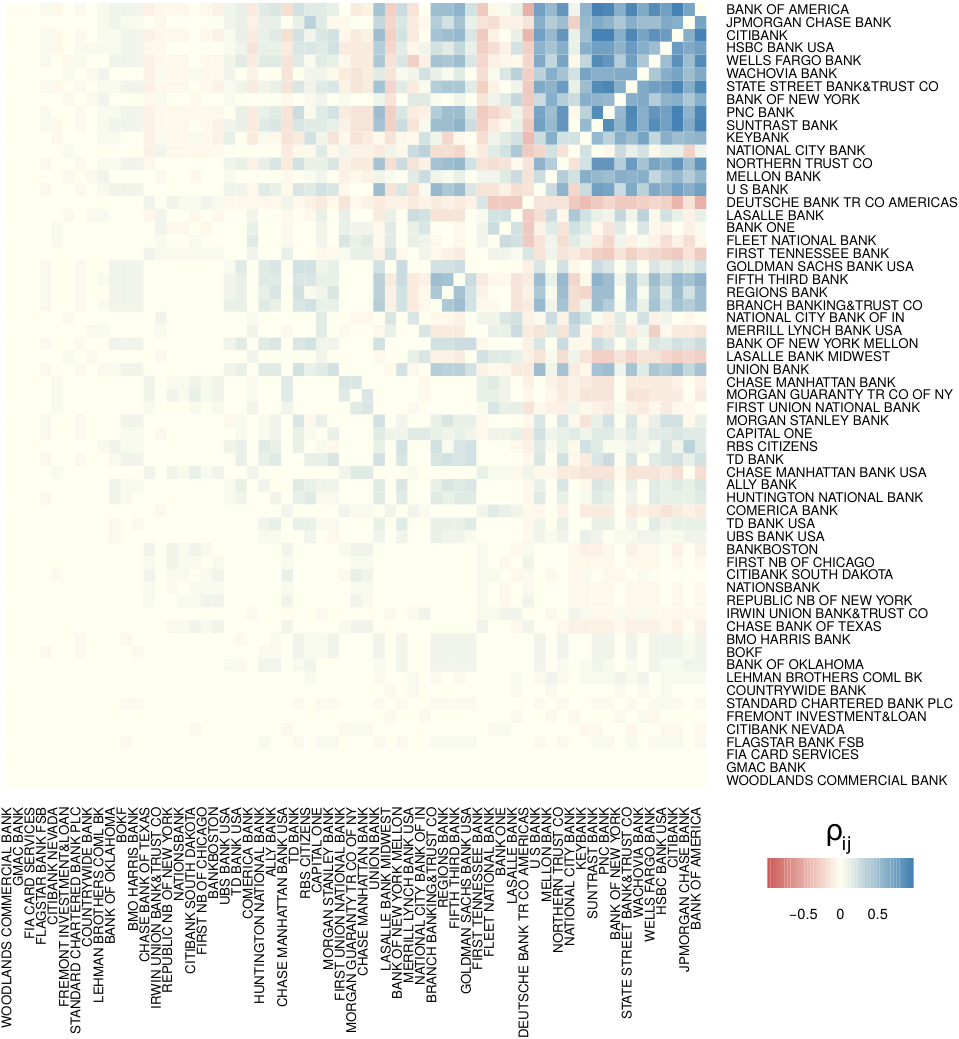}
  \end{center}
  \caption{non-normalized Pearson correlation coefficients $\rho_{ij}$ (based on pairwise available data), scaled by $w_{ij}$ with $l_{ij}=1$.}
  \label{fig:corrfin}
\end{figure}

The resulting correlation matrix is shown in Fig.~\ref{fig:corrfin}. Compared to the non-scaled Fig.~E in \nameref{S1_Appendix}, both the correlated and the anti-correlated activities loose importance for institutions with higher ranks, because the co-appearance in the ranking is rather sparse. But still, it is obvious that the correlated activities are concentrated in the core, while the anti-correlated activities can be mostly found in the periphery. Keeping in mind the exponential growth of the derivative volume of some of the key players, as shown in Fig.~\ref{fig:corrcomp}, it means that the OTC market acted rather \emph{heterogeneous}. Most banks with high ranks, i.e., key players, increased their activities in a growing market. Banks with lower ranks, such as \textit{First National Bank of Chicago} or \textit{RBS Citizens}, have either reduced their overall OTC exposure or have concentrated their activities towards only mayor institutions, avoiding other low ranked institutions.  

\subsection*{Correlations in risk}
\label{sec:corr-risk}

So far, we have only analysed correlations in \emph{activities}, i.e., the correlated increase or decrease in OTC derivatives volumes between any two institutions. We found that the correlated behavior was the dominating one which, together with a vastly growing OTC market, implies that most institutions increased their involvement. The question remains what this would mean for the \emph{risk} of the counterparties, 

We already mentioned in Section \nameref{otc:risk} that credit risk is the main source of risk for banking institutions. 
To estimate the \emph{total credit exposure} (TCE), we sum up their \emph{current credit exposure} (CCE) and their \emph{potential future exposure} (PFE) as explained in Section \nameref{otc:risk}. 
This data has been made available in ``Table 4'' of the OCC reports for each quarter year (see Table~B in \nameref{S1_Appendix}) and is used for our subsequent correlation analysis. ``Table 4'' lists, for each of the 25 first ranked institutions, the \emph{bilaterally netted current credit exposure} and the \emph{bilaterally netted potential future exposure} and the sum of both, TCE=CCE+PFE, as reported by the institutions themselves. Looking at Q1 of 2012, we first notice that, for the high ranked institutions (according to their activity in OTC derivatives), the potential future exposure exceeds considerably the current exposure, which is generally not the case for the lower ranked institutions. The question whether this observation is related to the financial crisis of 2008 is addressed further below. 

\begin{figure}[htbp]
  \centering
  \subfigure[Bilaterally netted current credit exposures]{
    \includegraphics[width=0.49\linewidth]{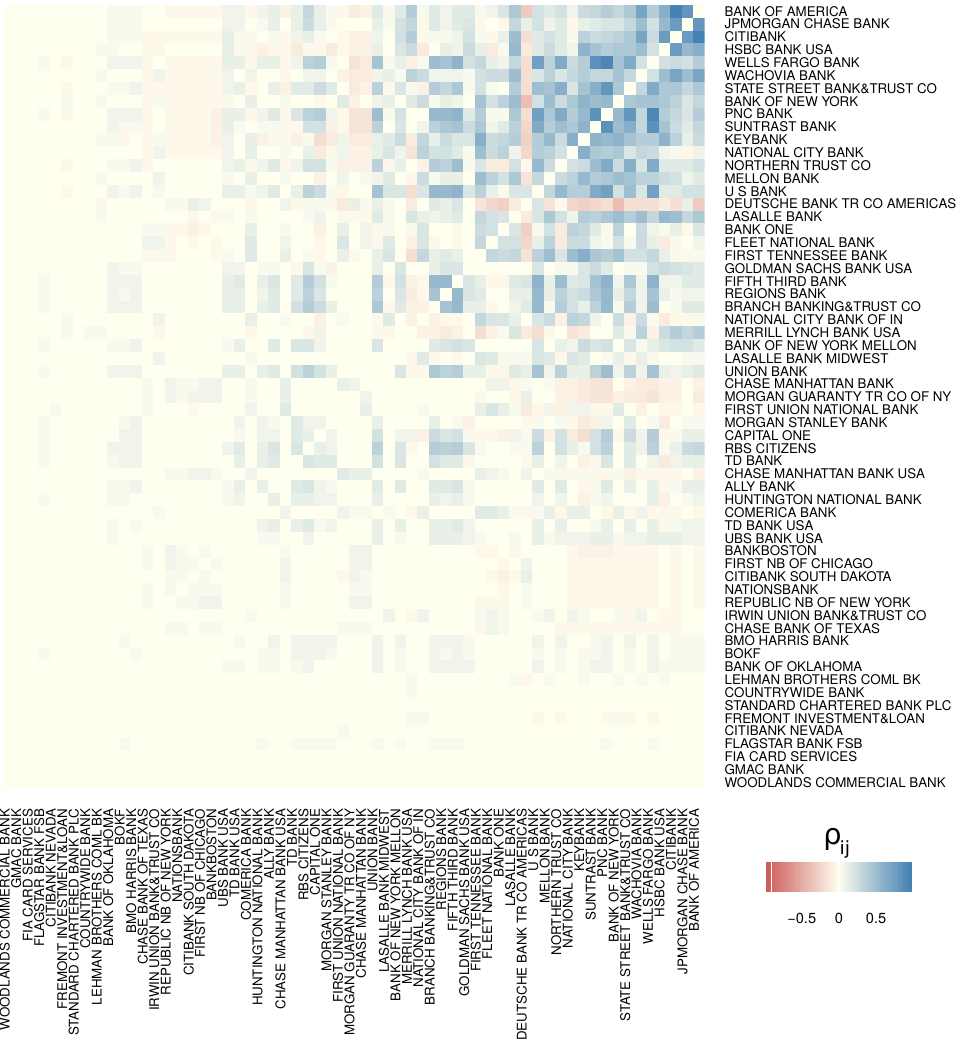}
    \label{fig:cce}
  }\subfigure[Total credit exposure]{
    \includegraphics[width=0.49\linewidth]{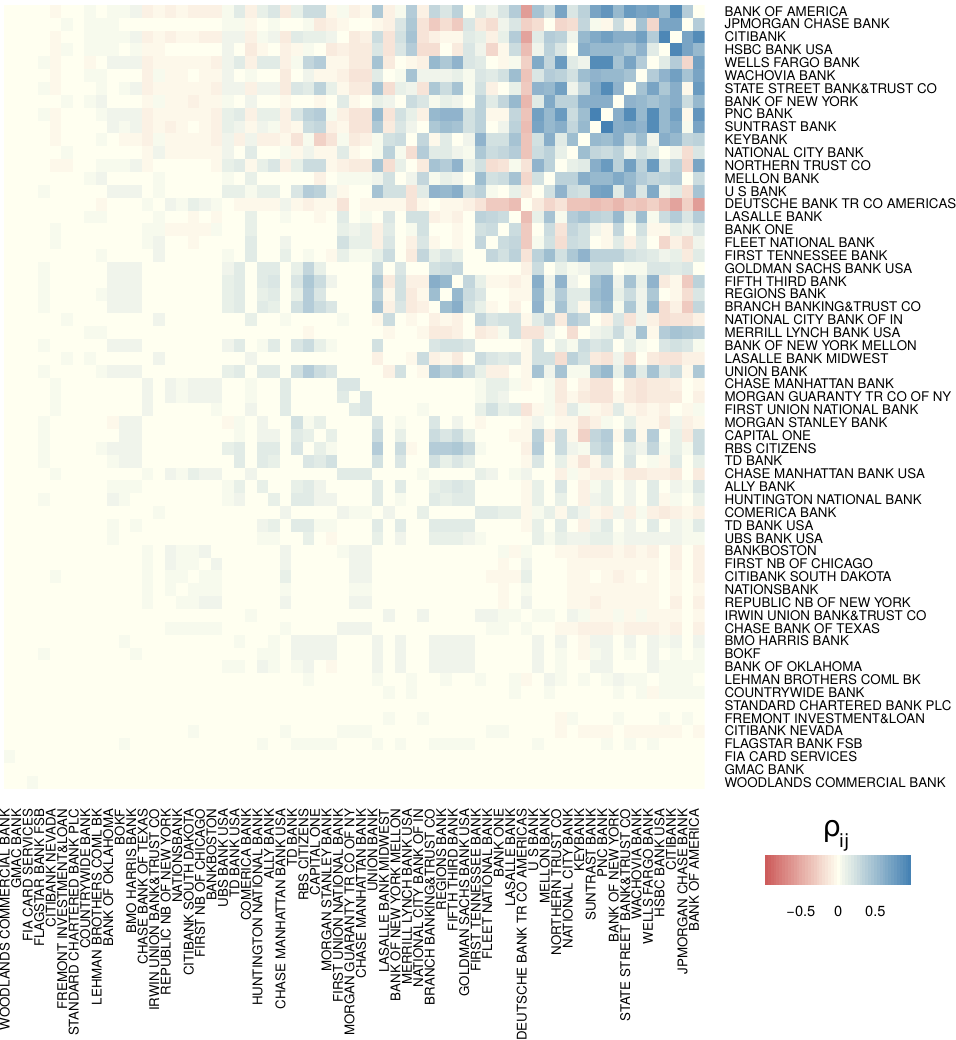}
    \label{fig:tce}
  }
  \caption{Non-normalized Pearson correlation coefficients $\rho_{ij}^{\mathrm{CCE}}$ and  $\rho_{ij}^{\mathrm{TCE}}$ (based on pairwise available data), scaled by $w_{ij}$ with $l_{ij}=1$.}
  \label{fig:correxposure}
\end{figure}
We can now define a correlation coefficient $\rho_{ij}^{\mathrm{TCE}}$ based on Eq.~(\ref{eq:pearsonT}) by just replacing the values of the activities $a_{i}(t)$ by $\mathrm{TCE}_{i}(t)$, and for $\rho_{ij}^{\mathrm{CCE}}$ accordingly. Following the argumentation above, we weight these correlations again by the weights $w_{ij}$. The results are shown in Fig.~\ref{fig:correxposure}. 
Both figure parts indicate that, at least for the  subset of banks which are the closest to the core according to the core-periphery decomposition, 
the credit exposures are highly positively correlated.
This indicates that the core of the network consists of institutions which are very strongly interdependent. This can become a reason for systemic instability, as the credit exposures and the connected risks  cannot be well diversified. 

The correlation pattern for the risk resembles the one found for the activities, Fig.~\ref{fig:corrfin}.
We have to note, however, that a large correlation coefficient $\rho^a_{ij}$  is a good indicator of a long-term activity between institutions $i$ and $j$, but a large correlation coefficient $\rho_{ij}^{\mathrm{TCE}}$ does not allow us to derive such a conclusion.

Up to this point the analysis was based on the whole available period of time (1998 -- 2012). It is interesting to repeat the correlation analysis of risk for the time before and after the financial crisis, separately. We avoid to discuss the precise mapping of ``before'' and ``after'' and have chosen Q4 of 2008 to divide  the  time series  into two periods. In Q4 of 2008 \textit{Goldman Sachs} entered the ranking of the OCC, for the first time, right after the collapse of \textit{Lehman Brothers} on 15 September, 2008. 

\begin{figure}[htbp]
  \centering
  \subfigure[Total credit exposure before 2008 Q4]{    \includegraphics[width=0.45\textwidth]{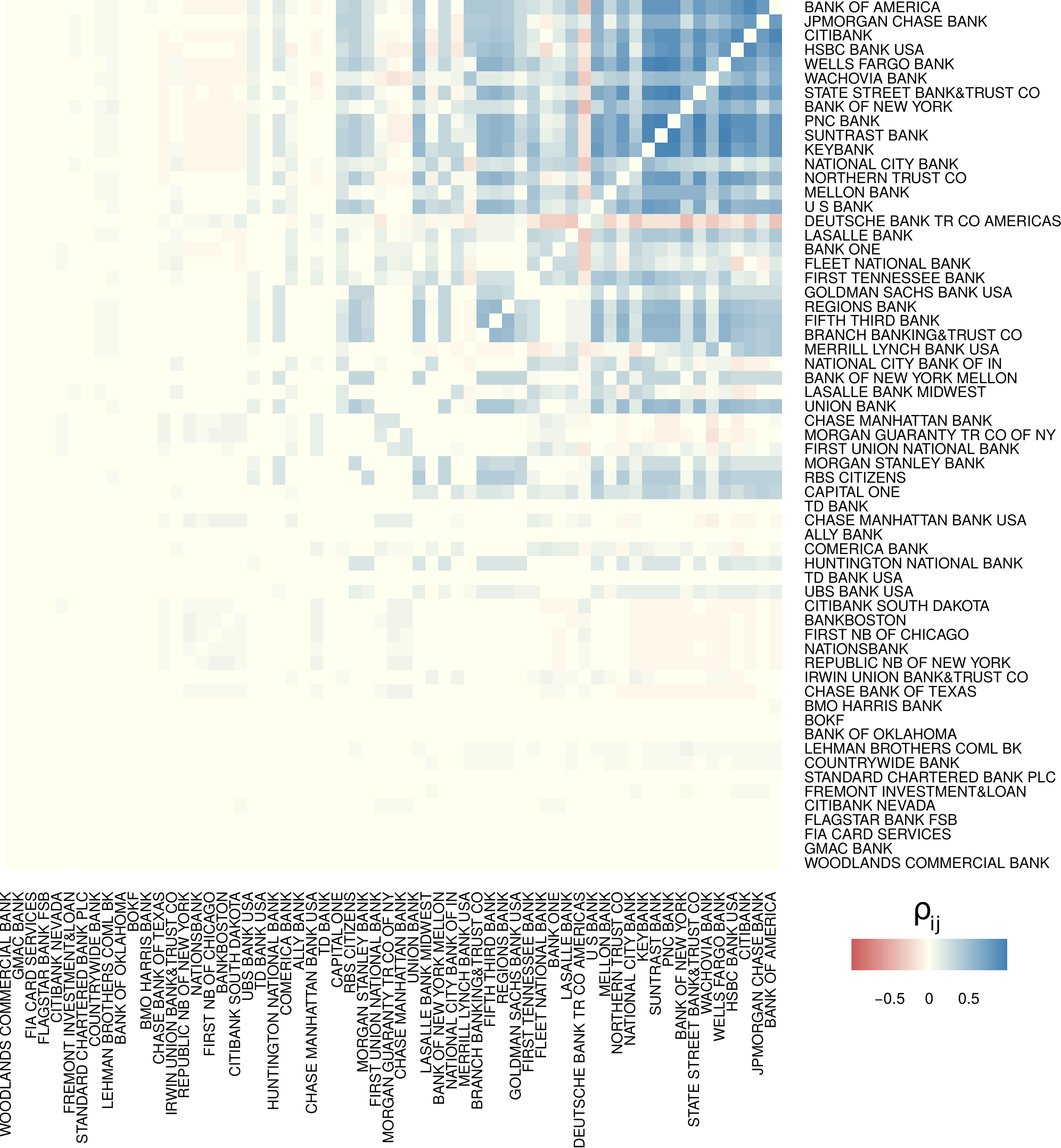}\label{fig:tcebefore}
  }\subfigure[Total credit exposure after 2008 Q4]{\includegraphics[width=0.45\textwidth]{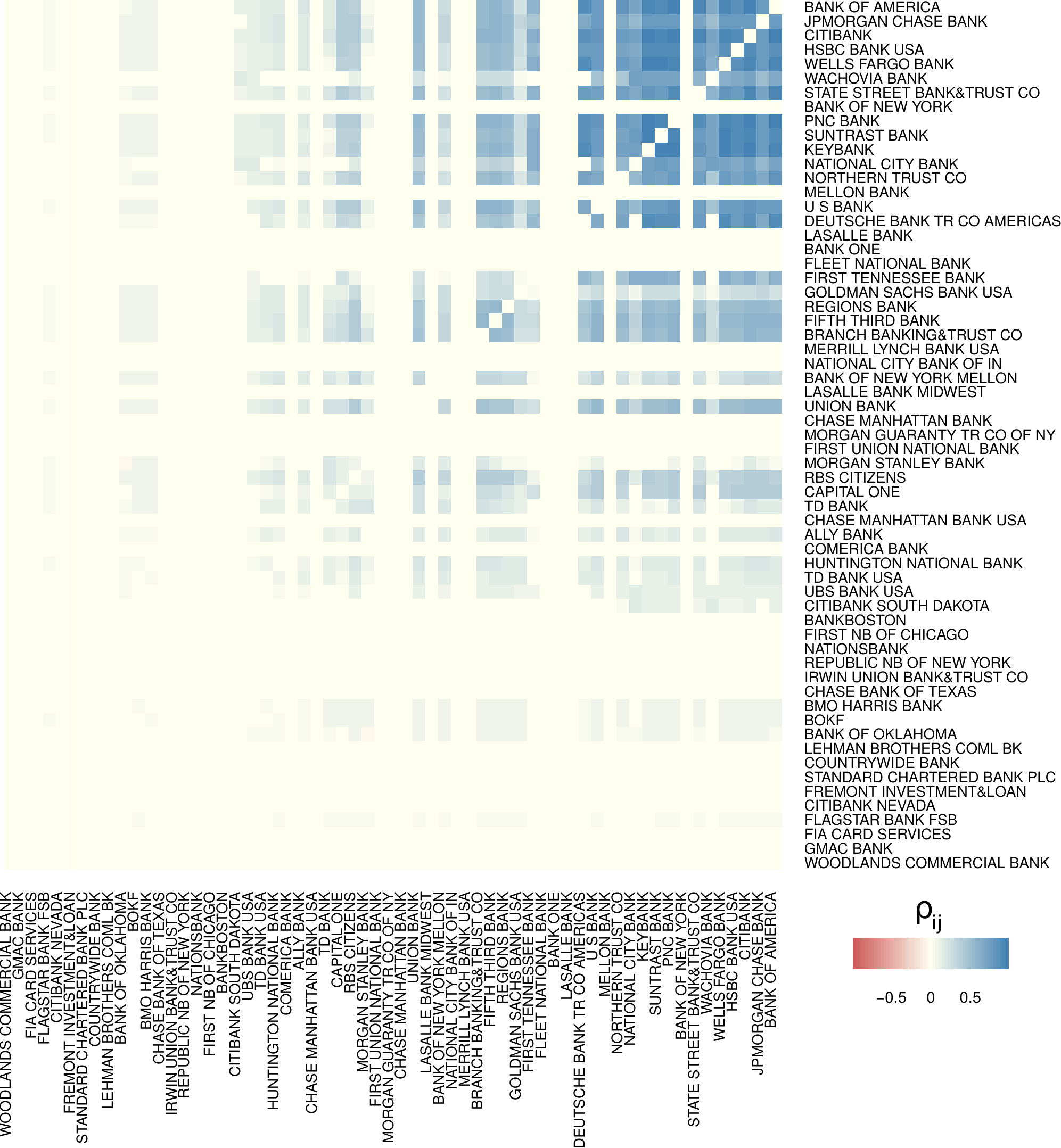}\label{fig:tceafter}
  }
  \caption{Correlations in total credit exposure (a) before and (b) after the financial crisis of 2008 Q4.}
  \label{fig:crisis}
\end{figure}
The results of our analysis before and after Q4 of 2008  are shown in Fig.~\ref{fig:crisis}. 
Comparing the two parts of the figure, we make two observations: 
(i) All listed banks follow a similar  behavior before and after the crisis.  But after the crisis the correlations became  more homogeneous and non-negative even  between  low-to-low ranked and low-to-high ranked institutions. 
(ii) Except only few banks, the key players in the core did not change.
Therefore, the OTC derivatives market structurally remained the same despite its vast growth.

\section*{Conclusions}
\label{sec:conclusions}

Our investigation reveals the \emph{hidden} network structure behind the OTC market in the United States, and the network evolution from 1998 to 2012. 
For this, we use publicly available data from the \cite{OCC} reports, which contains aggregated numbers about the activities of financial institutions, measured by the volume of their different derivatives.

We focus on two different aspects: (i) \emph{co-occurrence patterns} of institutions, which take into account their ranks and activities to reconstruct the network of counterparty risk. 
This network was further analysed using of a weighted k-core method, to reveal its core-periphery structure. 
This allowed us to compare the topology-based ranking with the activity-based ranking, and to identify the most important institutions and their mutual relations. 
(ii) \emph{correlation patters}, to reveal \emph{dependencies} in activities, and the subsequent counterparty risks of any two institutions. 
Our findings, namely an emergence of a pronounced core and the higher correlations in credit exposure associated with it, hint at \emph{increasing} counterparty and systemic risk in OTC derivatives market.

One could argue that the list of the few top institutions with the highest counterparty risk is not really surprising and financial experts would have known this anyway. 
But the point of our investigation is to present a formal, yet simple approach, to \emph{decompose} their \emph{known aggregated activities} into \emph{unknown bilateral exposures}. 
Only this allows us to reveal the hidden \emph{network}, and to estimate the \emph{systemic risk}.
Counterparty risk is not just the sum of individual risks, but can be amplified over the network of dependencies.
Precisely, the failure of single institutions, even in the periphery of the OTC network, can lead to the collapse of the whole system because of distress and load distributed over the network \cite{Garas2015}.

Such considerations do not only enhance our understanding of systemic risk, they also allow to develop more refined risk measures, and a more realistic pricing of OTC contracts. 
This network perspective is missing in existing investigations \cite{Miguel2008,Singh2010} on systemic risk in OTC derivatives markets. 
It moves the focus from discussing netting procedures \emph{after} a default to the more important question of how systemic risk \emph{emerges}, i.e., what happens \emph{before} a default.

To conclude, our investigations contribute to the ongoing debate about the impact of the OTC derivatives market on the stability of the financial markets. 
We support the position that OTC derivatives increase financial instability, because they generate a hidden network of dependencies that at the end increase the chance of failure cascades.  
This has not become obvious because of the bilateral nature of counterparty risk and the lack of transparency in OTC markets. 
But our simple and practical method allows to at least estimate this hidden network of additional dependencies and to better estimate, and price, the risk resulting from these.
It particularly points to the limitations in diversifying risk in such markets and the need to implement further regulations, as already proposed by the European Market Infrastructure Regulation (EMIR) under the Basel III
umbrella BCBS (2011).

\section*{Acknowledgments}

The authors thank Stefano Battiston and Paolo Tasca for discussions, and Michelangelo Puliga for pointing us to the dataset.
We acknowledge financial support from the 
Project CR12I1\_127000 \emph{OTC Derivatives and Systemic Risk in Financial Networks} financed by the Swiss National Science Foundation.

\clearpage

\section*{Supporting Information}
\label{S1_Appendix}

\renewcommand{\thetable}{\Alph{table}}
\renewcommand{\thefigure}{\Alph{figure}}
\setcounter{figure}{0}
\setcounter{table}{0}

\subsection*{Appendix 1: Data Availability and Processing}
\label{sec:app:0}

Our quantitative analysis is based on a dataset derived from the  \emph{quarterly} reports on derivatives of the Office of the Comptroller of the Currency (OCC). 

These reports contain, for each quarter, different tables with derivatives data of the top 25 US national insured commercial banks and trust companies. 
 A typical ``Table 1'' from the first quarter of 2012 \cite{OCC} is shown in Tab.~\ref{tab:t1} in the Appendix. 
The unit for all numbers is 1 million US dollars. 

The first column contains the ranks and the second column the name of the  institution. 
The data used for the ranking can be found in the fifth column, labeled \emph{Total derivatives}. It gives a proxy for the \emph{activity} of an institution.  The most active one  is assigned rank 1 and referred to as the ``highest ranked'' (or best ranked) in the following. In the second set of columns, ``Table 1'' of these reports further presents the \emph{composition} of the derivatives contracts into \emph{exchange traded} derivatives (ETD) (futures contracts, option contracts) and \emph{Over-The-Counter} (OTC) traded derivatives (forwards, swaps, options and credit derivatives). In the last column, foreign exchange spots are reported but not included in the sum of total derivatives. 

Below the list of the ranked institutions, Tab.~\ref{tab:t1} reports three important rows: 
(a) the sum of the derivative contracts for \emph{the top 25 ranked} institutions, (b)  the sum of the derivative contracts of \emph{all other} reporting institutions  (with the number of these stated in the reports until 2007/Q1 and (c) the sum of (a) and (b) referring to the whole market.

Such Tables are available for each quarter between  1998/Q4 and 2012/Q4, i.e., for 57 quarters.

The full OCC quarterly reports on derivatives are available on the OCC website \cite{OCC} in PDF and raw data is available in XML format. 
In order to obtain the data on credit exposures from derivative contracts, the latter was chosen and processed with the \emph{XML} package of \emph{R} statistical environment. 
A technical problem was met and  solved at this stage. The XML parser would not process random rows of the data. 
Inconsistencies in the structure of XML were the source of the problem and were found by manually checking the problematic rows. In the these rows one or two excessive empty data nodes were present, after deletion of which the whole available data was parsed. 
No XML file is published by OCC for the 2008 Q3, so the PDF file of the report was processed to obtain data and add it the rest of the dataset.

The next step after obtaining the data on credit exposures was merging it with the original dataset on notional amounts of derivative contracts. 
A technical hurdle was in matching different typesetting of the names of some institutions (e.g. extra spaces). 
At this stage few missing entries in the original dataset were found and fixed.

The raw data was combined into a CSV table and included 17 columns with the data from the mentioned two tables of the OCC reports. 
The chosen time range corresponded to 57 quarters, with 25 entries for every one summing up to 1425 rows in the table. 
Every row included the time-stamp, name of the institution, its rank, state, total assets, total derivatives, total amounts of 7 types of derivatives, bilaterally netted current credit exposure, potential future exposure, total credit exposure from all contracts and total credit exposure to capital ratio (the last in percents). 

The number of unique combinations of name and state in the raw dataset was 93, but the states were left out of consideration. 
Thus in the present analysis two actors with the same name but from different states were considered to be the same actor. 
The technical legitimacy of this assumption is based on the fact that no two actors with the same name but from different states simultaneously appeared in the reports. 
Checking the legal validity of the assumption is not in the scope of this analysis. 

The number of entries differing by the name of the institution was 82 in the raw data. 
The names were processed and their number reduced to 61. 
The conditions for merging two similar names were the absence of simultaneous appearance in the reports, similar ranks and orders of reported numbers. 
Again, legal aspects for merging these data were not considered. 
The main source of ambiguity in names were ``NA'' and ``NATIONAL ASSN'' endings. 
These were dropped and so, for example, ``KEYBANK NA'' and ``KEYBANK NATIONAL ASSN'' became ``KEYBANK''. 
Other more specific changes in the names are presented in the list below (in bold is the chosen name).  

\begin{description} {\scriptsize
  \item[CAPITAL ONE] \hfill \\
  Names in the original data were CAPITAL ONE BANK and CAPITAL ONE NATIONAL ASSN
  \item[DEUTSCHE BANK TR CO AMERICAS] \hfill \\
  BANKERS TRUST CO was renamed to DEUTSCHE BANK TR CO AMERICAS in April of 2002 
  \item[MELLON BANK] \hfill \\
  MELLON was misspelled MELLONG
  \item[SUNTRUST BANK] \hfill \\
  The bank was reporting as SUNTRUST BANK ATLANTA before 2000
  \item[UNION BANK] \hfill \\
  The bank was reporting as UNION BANK OF CALIFORNIA until the 3\textsuperscript{rd} quarter of 2008
  \item[BANKBOSTON] \hfill \\
  In the report from 1998 Q4 the name was BANKBOSTON CORPORATION
  \item[BANK OF AMERICA] \hfill \\
  BANK OF AMERICA was BANK OF AMERICA NT\&SA until the mid 1999.
  \item[FIRST TENNESSEE] \hfill \\
  The bank was misspelled as FIRST TENESSE in one entry
  \item[BMO HARRIS BANK] \hfill \\
  The bank appeared as HARRIS TRUST\&SAVINGS BANK
  \item[LASALLE BANK MIDWEST] \hfill \\
  STANDARD FEDERAL BANK changed its name after being acquired by LaSalle Corp.\ in 2005 
}
\end{description}

As described above, some banks which changed their names in the considered time period were given the latest name for the whole time period, as in the present analysis the institutions are distinguished by name. 
Assigning unique ID numbers to the institutions would make the presentation of the results more abstract and less understandable.

\FloatBarrier
\clearpage

\subsection*{Appendix 2: Activities and ranks}
\label{app:A}

\begin{figure}[htbp]
  \centering
  \includegraphics[width=0.38\textwidth]{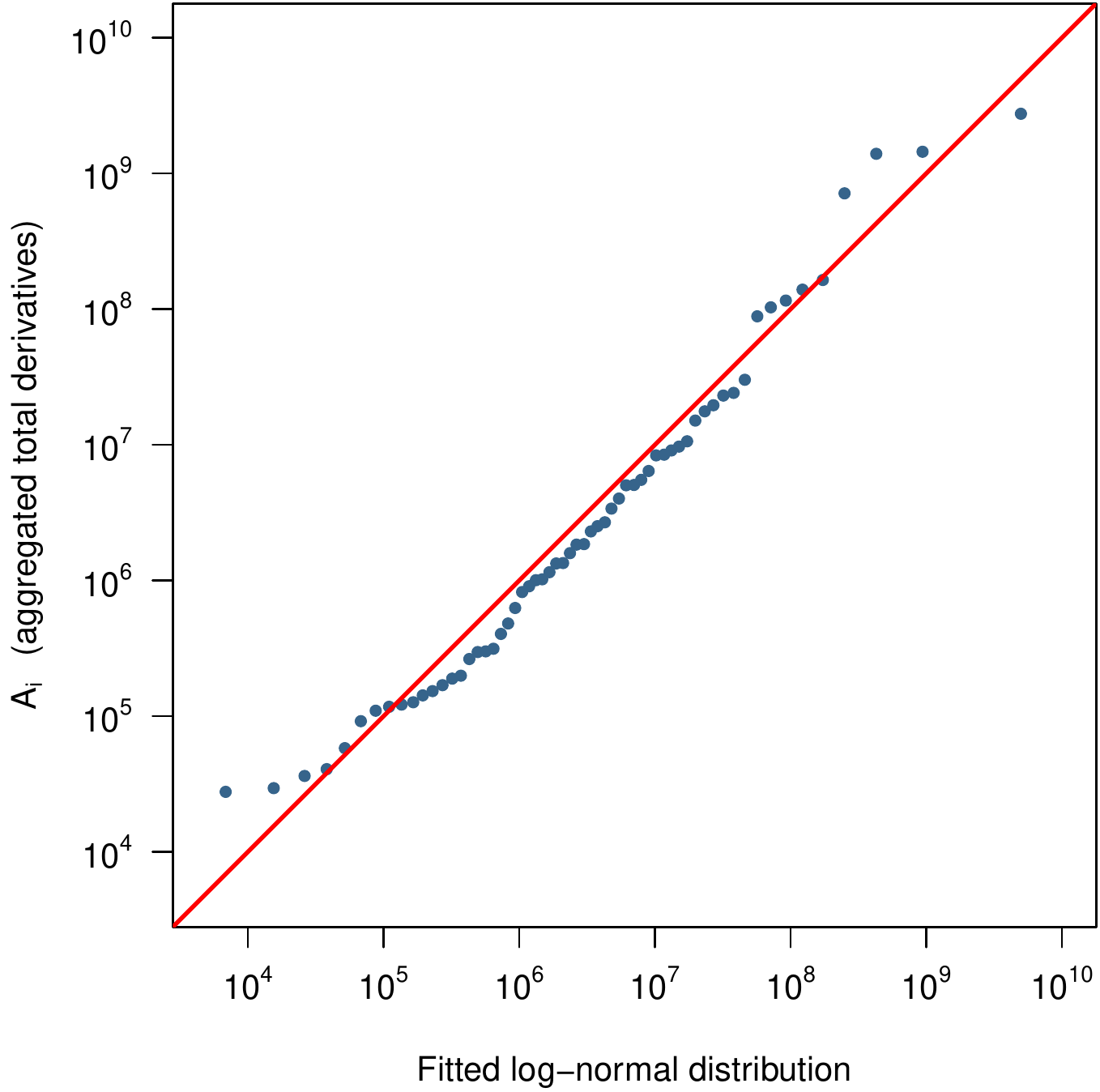}
  \caption{Quantile-quantile plot of the aggregated activity $A_{i}$ versus the fitted log-normal distribution, $\mu$=14.54116, $\sigma$=2.865165. In order to check if log-normal distribution is a good candidate to describe the aggregated total derivatives distribution, 10000 Kolmogorov-Smirnov two-sample tests were made for $A_i$ against synthetic samples, from which 9856 tests were positive for $p$-value equal to  0.10.}
  \label{fig:qq}
\end{figure}

 \begin{figure}[htbp]
  \centering
  \includegraphics[width=0.4\textwidth]{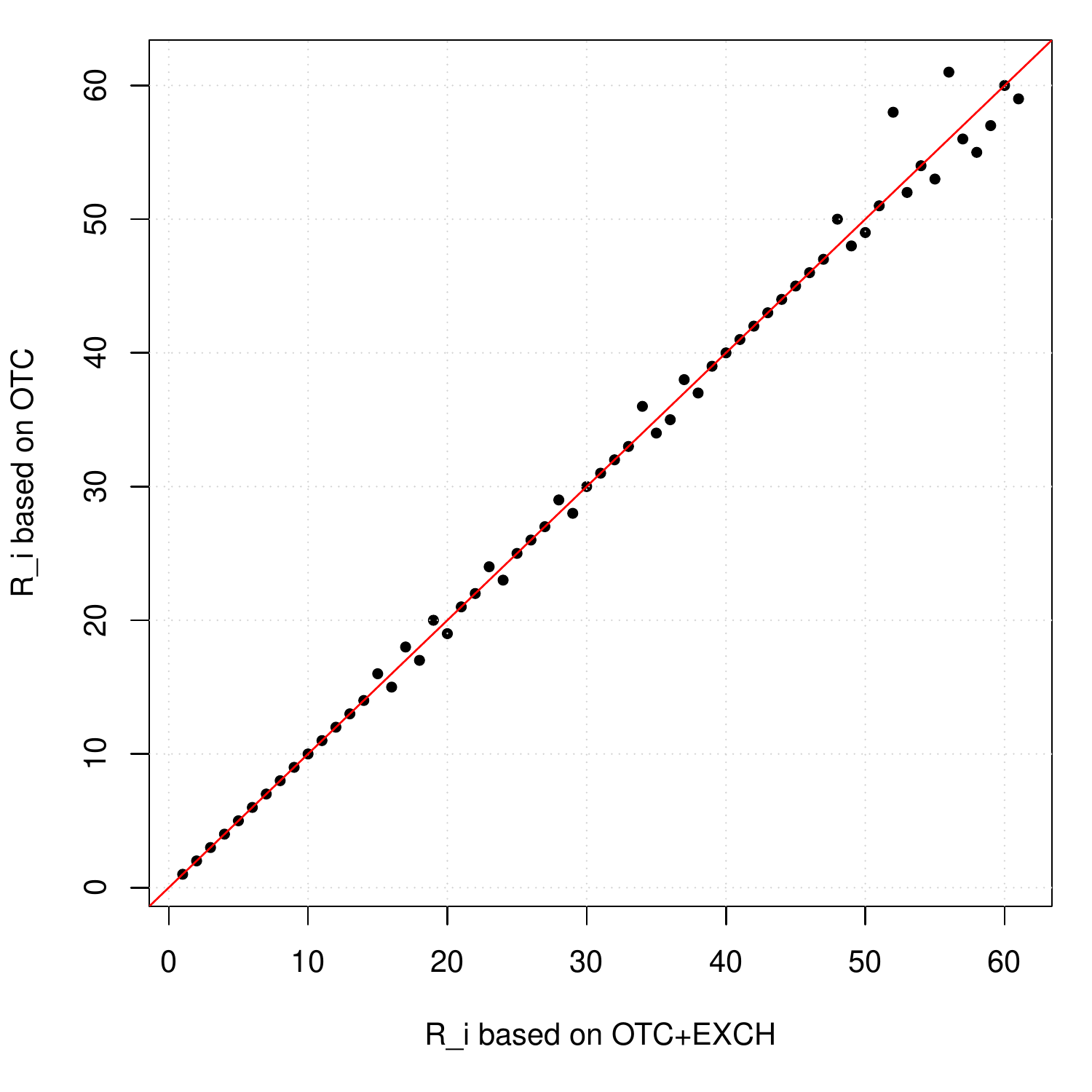}
  \caption{Ranks $R_{i}$ based on the total activity $A_{i}$ versus Ranks $R_{i}^{\mathrm{OTC}}$ based on the activity resulting from OTC derivatives $A_{i}^{\mathrm{OTC}}$}
  \label{fig:r-i}
\end{figure}

\begin{figure}[htbp]
  \begin{center}
     \includegraphics[width=.7\linewidth]{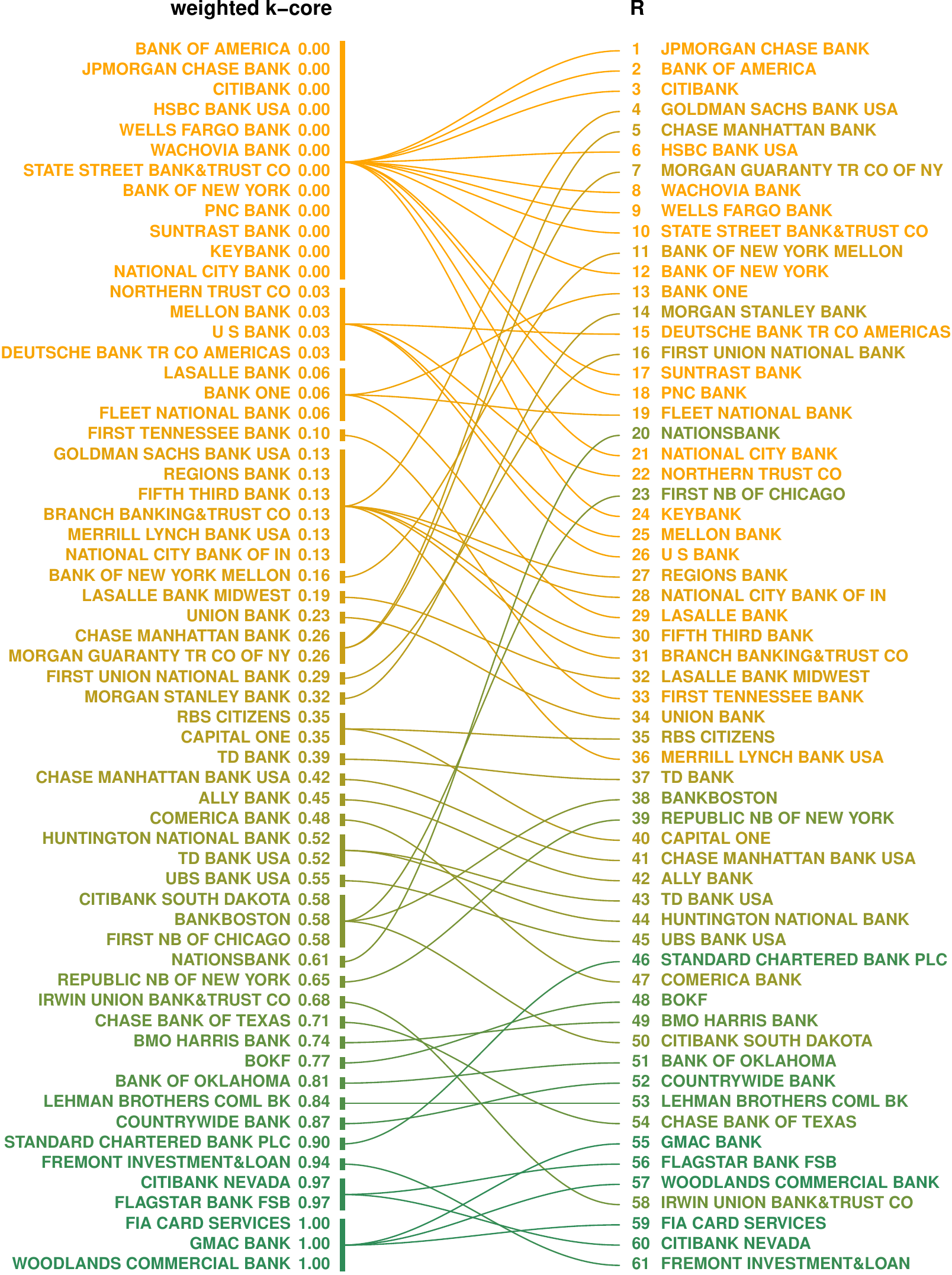}
  \end{center}
  \caption{Comparison of the weighted $k$-core (left column) and $R_i$ (right column) rankings. Better ranks are on the top. Links connect the same institution in two rankings. The colors represent $k$-core ranking, orange changes to green as the distance from core increases.}
  \label{fig:corrPearsonweighted}
\end{figure}

\FloatBarrier
\clearpage

\subsection*{Appendix 3: OTC network evolution}
\label{app:B}

\begin{figure}[htbp]
  \begin{center}
   \animategraphics[controls,width=0.7\linewidth]{2}{temp_net_corr}{0}{56}
  \end{center}
  \caption{Animation showing the evolution of the financial institutions' network over time. An institution is present in the network at certain time step if it is in the top 25 commercial banks, savings associations or trust companies in derivatives. The network of these 25 institutions is considered fully connected, with the weight of a link being proportional to the inverse of the lowest rank of its end nodes. The size and color of a node represents the significance of the node in terms of the sums of the weights of its links.}
  \label{fig:tempnet}
\end{figure}

\FloatBarrier
\clearpage

\subsection*{Appendix 4: Correlations}
\label{app:C}

\begin{figure}[htbp]
  \begin{center}
     \includegraphics[width=.75\linewidth]{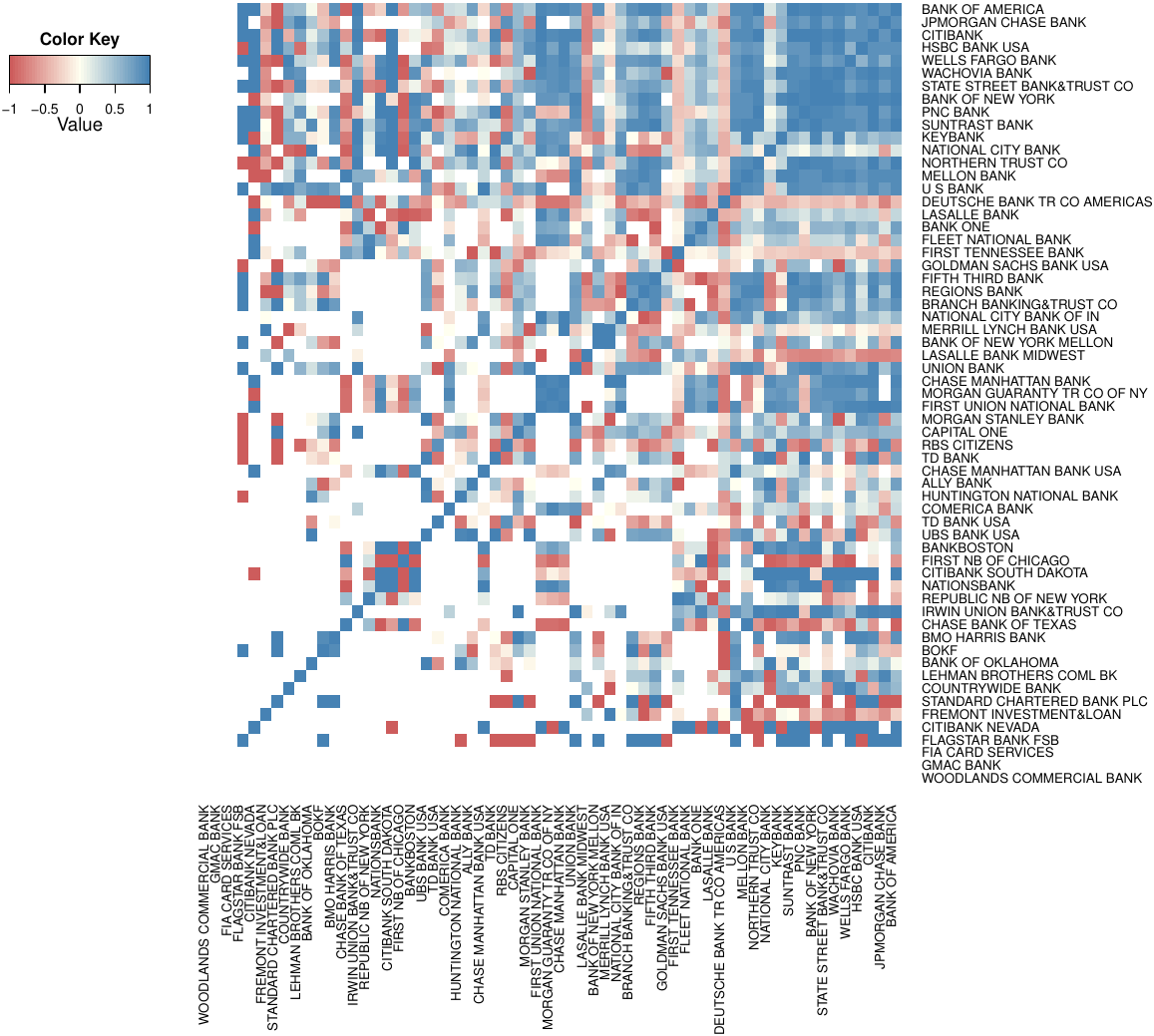}
  \end{center}
  \caption{Based on only pairwise available observations.}
  \label{fig:corrall}
\end{figure}

\subsection*{Appendix 5: Tables}
\label{app:D}

\FloatBarrier

\begin{sidewaystable}[htbp]
\caption{``Table 1'' from the OCC report for the first quarter of 2012.}
  \centering
  \begin{tabular}{c}          
      \includegraphics[width=0.99\textwidth,clip=true, trim=2.5cm 2.5cm 2.5cm 1.5cm]{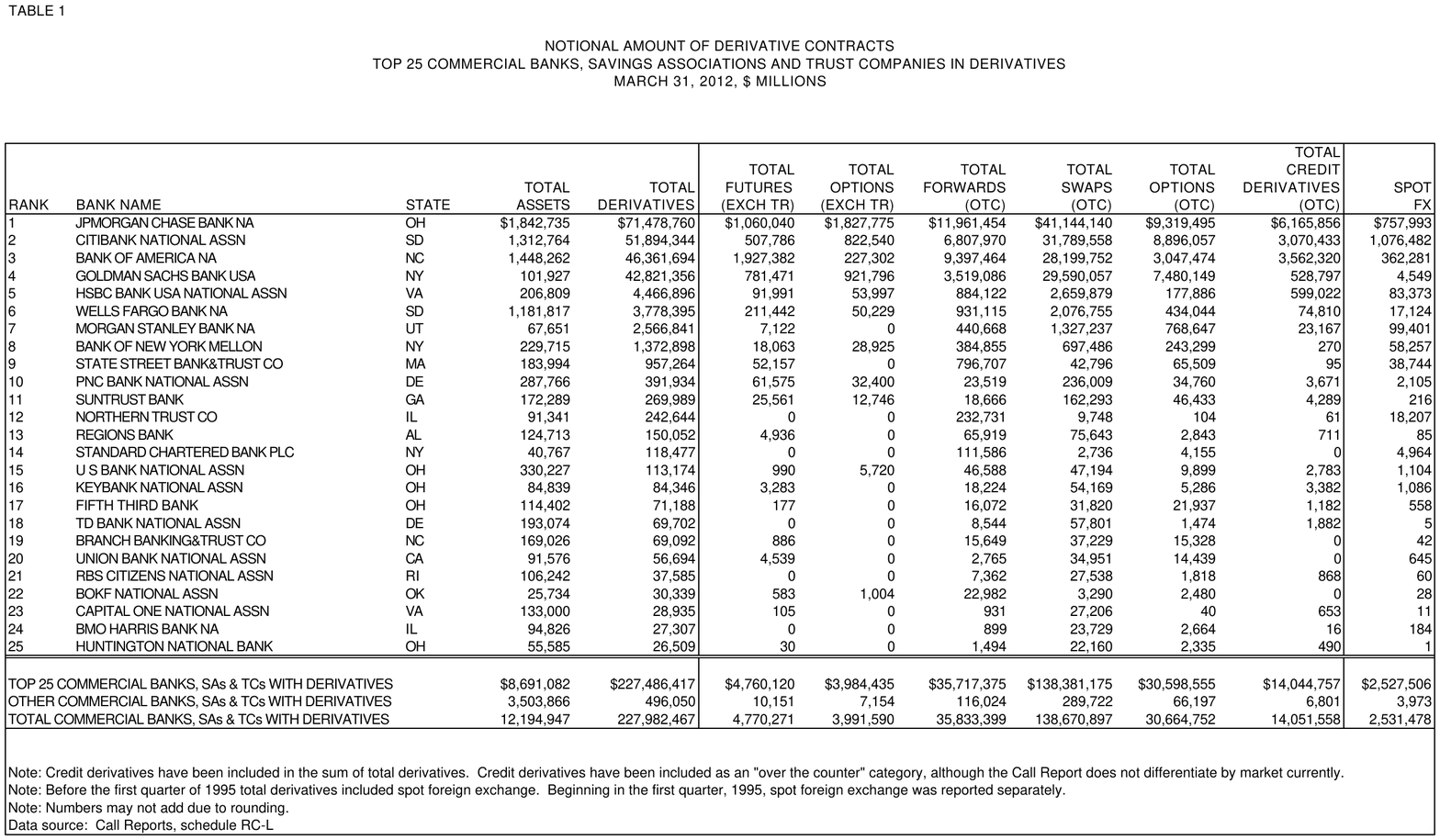} 
  \end{tabular}
\label{tab:t1}
\end{sidewaystable}

\begin{sidewaystable}[htbp]
  \caption{``Table 4'' from the OCC report for the first quarter of 2012.}
  \centering
  \begin{tabular}{c}
      \includegraphics[width=\textwidth, clip=true, trim=2.5cm 1cm 2.5cm 1.5cm]{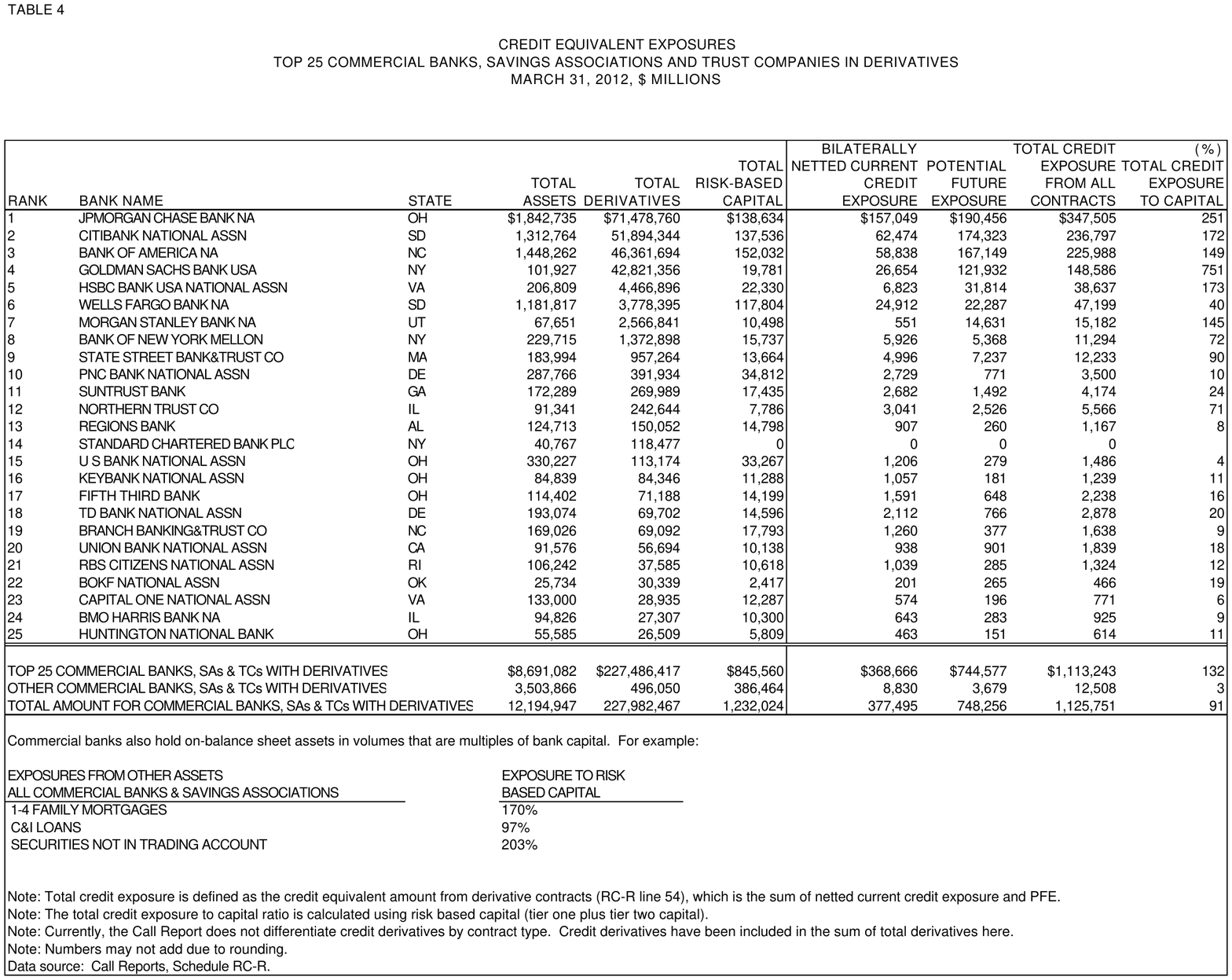}      
  \end{tabular}
  \label{tab:t4}
\end{sidewaystable}


\end{document}